\newcommand{\GG}[1]{\textcolor{black}{#1}}
\newcommand{\G}[1]{\textcolor{black}{#1}}

\documentclass[]{aa}
\usepackage{graphicx,url,twoopt,natbib}
\usepackage[varg]{txfonts}
\usepackage{subfigure} 
\usepackage{enumerate}
\usepackage{rotating} 

\bibpunct{(}{)}{;}{a}{}{,} 
\usepackage[colorlinks,
            linkcolor=red,
            anchorcolor=blue,
            citecolor=blue]{hyperref}
\graphicspath{{./}{figures/}}

\begin{document}

\title{The origin of B-type runaway stars based on kinematics}
\author{{Yanjun Guo}\inst{1,2},
         Chao Liu\inst{3},
         ZhiCun Liu\inst{4},
         Chunyan Li\inst{5},
         Qida Li\inst{6},
         Kun Chen\inst{1,2},
         Zhanwen Han\inst{1,2}
          \and
          XueFei Chen\inst{1,2}
          }
\institute{
            $^1$ Yunnan observatories, Chinese Academy of Sciences, P.O. Box 110, Kunming, 650011, China; cxf@ynao.ac.cn\\
            $^2$ International Centre of Supernovae, Yunnan Key Laboratory, Kunming 650216, China; guoyanjun@ynao.ac.cn\\
            $^3$ CAS Key Laboratory of Optical Astronomy, National Astronomical Observatories, Chinese Academy of Sciences, Beijing, 100101, People’s Republic of China\\  
            $^4$ Department of Physics, Hebei Normal University, Shijiazhuang 050024, People's Republic of China\\
            $^5$ Shanghai NanHui High School, shanghai, 200135, People’s Republic of China\\
            $^6$ Department of Astronomy, China West Normal University, Nanchong, 637002, People’s Republic of China\\
             }

\abstract
   {Runaway stars depart their birthplaces with high peculiar velocities. Two mechanisms are commonly invoked to explain their origin, the binary supernova scenario (BSS) and the dynamical ejection scenario (DES).
   Investigating the kinematic properties of runaway stars is key to understanding their origins.
}
    {We intend to investigate the origins of \G{39} B-type runaway stars from LAMOST using orbital traceback analysis.}
    {From the catalog of LAMOST, we selected \G{39} B-type runaway stars and determined their spectral subtypes from key absorption lines.
    We then derived atmospheric parameters for each star using the Stellar Label Machine (SLAM), which is trained on TLUSTY synthetic spectra computed under the non-local thermodynamic equilibrium (NLTE) assumption.
    Using the derived atmospheric parameters as input, we estimated stellar masses and ages with a machine learning model trained on PARSEC evolutionary tracks. 
    We finally performed orbital traceback with \texttt{GALPY} to analyze their origins.
    }
    {Through orbital traceback, we find that \G{29} stars have trajectories entirely within the Galactic disk, whereas \G{10} \G{are disk-passing} yet still trace back to the disk. 
    Two stars have trajectories that intersect those of known clusters. Their orbits show similar morphologies in both the $X-Y$ and $R-Z$ planes, and their [M/H] values are comparable, suggesting possible cluster origins. 
    However, definitive confirmation will require additional evidence.
    In addition, the $V_{\rm Sp} - v\sin{i}$ plane shows that runaway stars with low peculiar space velocities but high $v\sin{i}$ remain on the Galactic disk, whereas those with high peculiar space velocities but low $v\sin{i}$ \G{pass through} the disk, possibly reflecting two distinct origins.} 
    {}


\titlerunning{The origin of \G{} B-type runaway stars} 
\authorrunning{Yanjun Guo et al.}        
\maketitle

\section{Introduction} \label{sec:intro}
Runaway stars rapidly depart from their natal clusters or associations, typically exhibiting peculiar velocities of about 30 $-$ 40 km $^{-1}$ \citep{Blaauw1954,Gies1986,Hoogerwerf2000}.
Most observed runaway stars are massive early-type stars, with over 30\% of O stars and approximately 5$-$10\% of B stars considered to be runaway stars \citep{1961Blaauw,1979Stone,2005Mdzinarishvili,2023gaianewrunaway}. 
\G{While most runaway studies have focused on OB stars, cooler runaway stars such as A/F-type candidates have also been reported in the Milky Way and in the 30 Doradus region of the LMC \citep{2018Maiz,AFrunaway}.}
Furthermore, more than half of the massive runaway stars exhibit enhanced surface helium abundances and rapid rotation \citep{1993Blaauw,Hoogerwerf2000}.

Two formation scenarios have been proposed to explain their origin.
One is the binary supernova scenario (BSS), first proposed by \cite{1957Zwicky} and later developed by \cite{1961Blaauw}, in which the more massive component of a binary first undergoes core collapse, leaving behind a compact object (a neutron star or black hole). 
If the supernova ejects more than half of the system's total mass or imparts a significant natal kick to the remnant, the surviving companion is likely to become a runaway star \citep{2019Renzo}.
Otherwise, if the system remains bound, the companion and compact object may form a runaway binary, potentially observable as a high-mass X-ray binary (HMXB) \citep{2000van}.
The other channel is the dynamical ejection scenario (DES; \citealt{1967Poveda}), in which early--type runaway stars are ejected from dense clusters or associations through gravitational interactions.
Among these ejection mechanisms, binary--binary interactions are more efficient than binary--single star encounters \citep{1983Hoffer}.
Both scenarios are believed to occur, but their relative efficiencies have been debated for the past few decades \citep{2022Sana,2025CK}.

Runaway stars formed via DES and BSS can be distinguished in several ways.
One common method involves tracing back the kinematic trajectories of runaway stars to their parent clusters, which is widely used to identify runaway stars originating from the DES \citep{Blaauw1954,Gies1986,2001Hoogerwerf}.
The chemical composition also provides valuable information. 
Although runaway stars from both formation scenarios typically retain their initial chemical abundances \citep{2017McEvoy}, supernova explosions in binary systems may result in $\alpha$ element enrichment in ejected stars \citep{2008Przybilla,2010Irrgang}.
In addition, DES runaways typically have higher space velocities than BSS ones, while runaways with high $v\sin{i}$ are thought to arise mainly from BSS \citep{2022Sana}.

The recent availability of extensive spectroscopic data from the Large Sky Area Multi-Object Fiber Spectroscopic Telescope (LAMOST)  has provided an opportunity to investigate the origin of runaway stars \citep{2012CuiXiangQun,2012ZhaoGang,2012DengLiCai,2020LiuChao}. 
Recently, \cite{2024ApJSGYJ} identified 547 runaway stars from LAMOST Data Release 8 (DR8) and analyzed their statistical properties.
Based on that catalog, \cite{2025CK} investigated their binary fractions and evaluated the relative contributions of the BSS and DES formation scenarios.
In this work, we perform orbital traceback analysis for \G{39} B-type runaway stars from the sample identified by \cite{2024ApJSGYJ} to investigate their origins.

The structure of this paper is as follows. 
We introduce the LAMOST data in Section~\ref{sec:data}. 
Section~\ref{sec:Method} presents the subclassification of the sample and the methods used to estimate its atmospheric and fundamental parameters
Kinematic analyses and the corresponding results are presented in Section~\ref{sec:Results}. Finally, our conclusions are summarized in Section~\ref{sec:Summary}.

\section{Data} \label{sec:data}
The Large Sky Area Multi-Object Fiber Spectroscopic Telescope (LAMOST) is a 4-meter quasi-meridian reflecting Schmidt telescope located at the Xinglong station of the National Astronomical Observatories, China. 
The focal plane of the telescope offers a 5-degree field of view and can accommodate up to 4,000 fibers \citep{2012CuiXiangQun,2012ZhaoGang,2012DengLiCai}.
Since 2012, LAMOST has conducted a Low-Resolution Spectroscopic (LRS) survey with a resolving power of $R\sim 1800$  and a wavelength coverage of $3690 \sim 9100$~\AA\ \citep{2012DengLiCai}.
In October 2018, LAMOST began a five-year Medium-Resolution Spectroscopic (MRS) survey aimed at conducting time-domain observations of objects in the Galactic region. 
The MRS spectra are captured with blue and red cameras, covering wavelength ranges of 4950 $\sim$ 5350 \AA\ and 6300 $\sim$ 6800 \AA\ , respectively \citep{2020LiuChao}. 

By combining LAMOST DR8 radial velocities with Gaia DR3 astrometry, \G{\citet{2024ApJSGYJ} determined the peculiar tangential, radial, and space velocities of 4,432 early-type stars.
Following the approach of \citet{Berger2001} and \citet{2022luqianBe},  we identified 229 runaway stars with peculiar space velocities exceeding 43 km s$^{-1}$, corresponding to approximately 1\% of the peak of the fitted Maxwellian distribution of the space velocities.
In addition, 480 further runaway candidates were selected based on their peculiar tangential velocities exceeding 33 km s$^{-1}$, corresponding to approximately 1\% of the peak of the fitted Maxwellian distribution of the tangential velocities.
Because 162 sources are common to both subsamples, the final catalog comprises 547 runaway stars.}

\begin{figure*}
    \centering
    \includegraphics[scale=0.6]{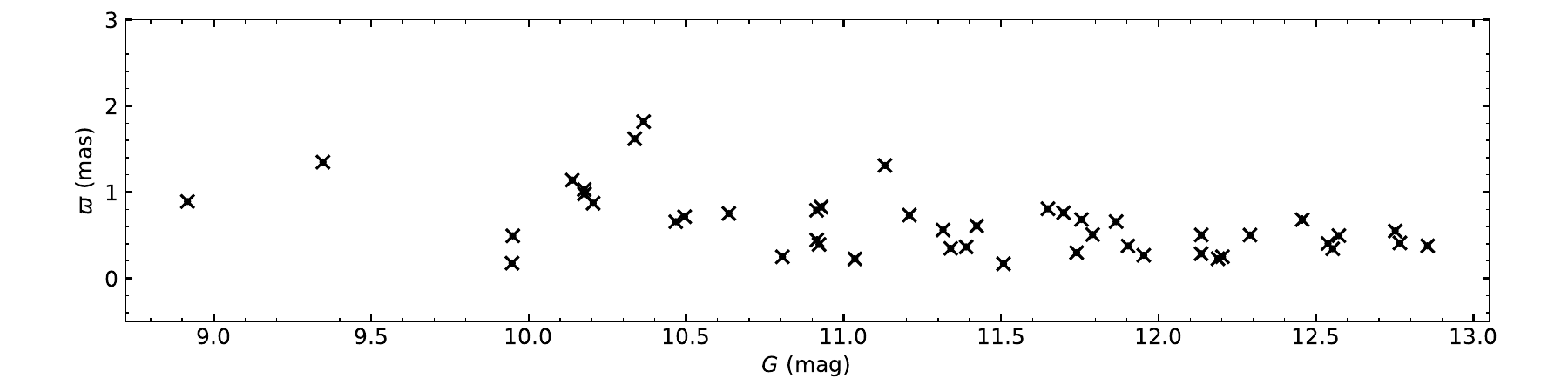}
    \caption{Gaia DR3 parallaxes as a function of $G$ magnitude. The error bars represent the parallax uncertainties.}
    \label{fig: Gmag}
\end{figure*}

From this catalog, we select 46 B-type runaway stars and determine their kinematic origins using orbital traceback.
\G{Although many studies restrict the definition of B stars to the B3–B9 range, we adopt the broader B0V–B9V range, since our sample also includes three B2-type stars.}
\G{The Gaia DR3 parallaxes as a function of $G$ magnitude for the 46 runaway stars are shown in Figure~\ref{fig: Gmag}.
Most objects are concentrated between $G = 10$ and $13$~mag.
Only four stars have $G < 10$~mag, two of which are at $G = 9.94$~mag, while the other two are somewhat brighter ($G = 9.3$ and $G = 8.9$~mag).
In general, the magnitude distribution of our sample is nearly homogeneous.}

\G{We note that some objects within the OB temperature range could, in principle, be contaminated by blue horizontal branch (BHB) stars, as both populations exhibit similar surface gravities \citep{2008Xue,2009Catelan,2013Salgado,2025BHBG}.
To assess this, we cross-matched our sample with the LAMOST BHB catalog \citep{2021Vickers,2024JuBHB} and several other major BHB catalogs \citep{2007Beers,2008Xue,2008Brown,2021Culpan}, and found one overlapping source.
This star has been removed from the final sample to avoid potential contamination, leaving 45 stars for further analysis.}

\section{Method} \label{sec:Method}
\begin{figure*}
    \centering
    \includegraphics[scale=0.5]{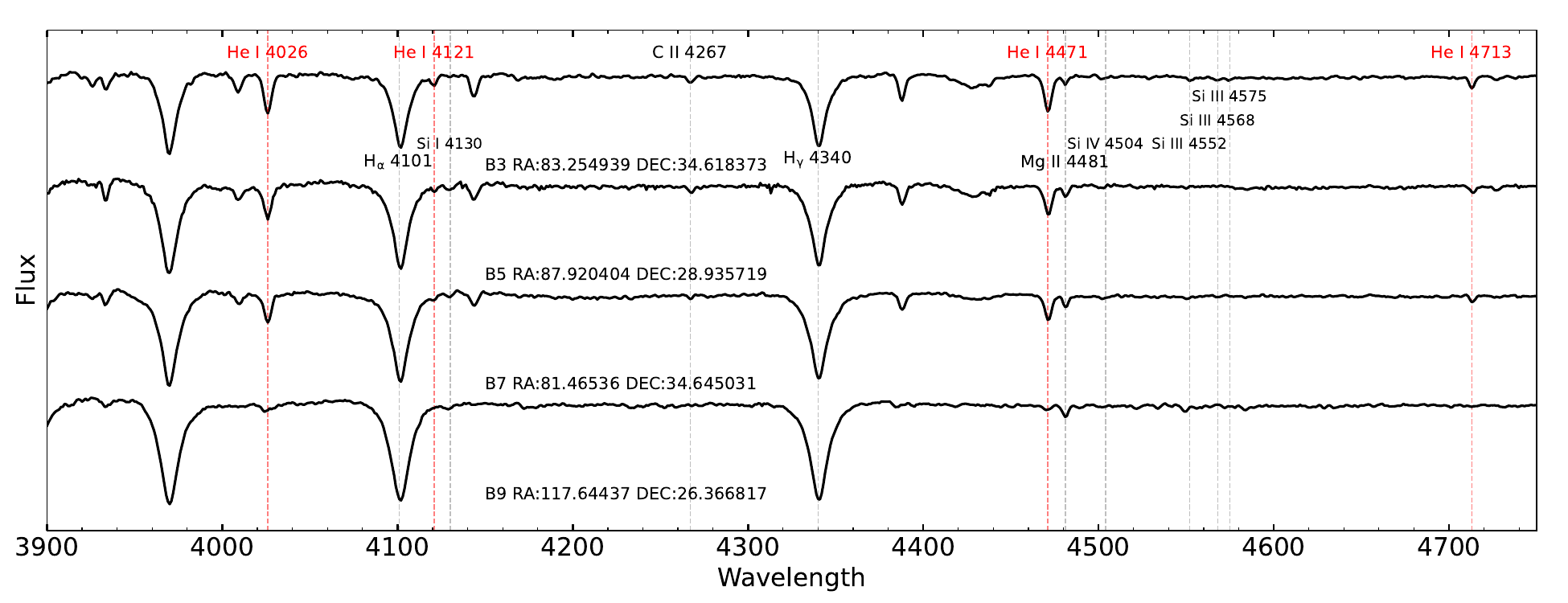}
    \caption{Sample spectra spanning B3, B5, B7, and B9 subtypes, arranged from top to bottom. Key spectral lines, including He I 4026, 4121, 4471, 4713, H$_{\delta}$ 4101, H$_{\gamma}$ 4340, and Mg II 4481, are labeled to highlight their variations across different subtypes. 
    The RA and Dec of each target are listed below the corresponding spectrum.}
    \label{fig: Type spectra}
\end{figure*}
\subsection{\G{Spectral classification}}
To obtain more precise atmospheric parameters, we cross-matched the 547 early-type star candidates with the LAMOST LRS DR10 and identified 123 matches. Because our analysis focuses on B-type runaway stars, we performed spectral subclassification of these spectra through visual inspection, following the MKCLASS-based criteria established for LAMOST OB stars by \citet{2019LiuZhicun} (see their Table 2).

\G{The classification criteria are summarized as follows: For B-type stars, the strengths of hydrogen Balmer lines (such as H$_\delta$ and H$_\gamma$) increase from spectral types B0 to B9. 
The neutral helium lines (\ion{He}{i}\,4026, 4121, 4471, 4713\,\AA) first strengthen and then weaken, reaching a maximum intensity at B2\,V. 
In contrast, the high-excitation \ion{Mg}{ii}\,4481\,\AA\,line gradually strengthens throughout the B-type sequence. 
When this trend is not clearly visible, we also rely on specific absorption lines and line ratios as auxiliary criteria.
For stars of types B2 to B9, the ratio of \ion{He}{i} 4471 to \ion{Mg}{ii} 4481 generally decreases \citep{2009Gray,2019LiuZhicun}.
Furthermore, for stars of the same spectral type, the strengths of the hydrogen Balmer lines generally decrease with luminosity from the main sequence to supergiants.}

Based on these classification criteria and the LAMOST-LRS spectra\footnote{ We used the python package {\tt{laspec}} from \url{https://github.com/hypergravity/laspec} to normalize the spectra using a spline fit.} from the LAMOST archive, we have determined the spectral types for \G{45} B-type runaway stars, which are listed in Column 9 of Table~\ref{Tab:1}. Figure~\ref{fig: Type spectra} shows the spectra of five representative B-type stars (B3 to B9) in the wavelength range of 3900–4750\,\AA, with some important spectral feature lines marked.

\subsection{\G{Atmospheric} parameters}\label{sec: Atmosphere parameter}
With the spectral subtypes established above, we derive atmospheric parameters for our B-type runaway stars, including effective temperature ($T_\mathrm{eff}$), surface gravity ($\log{g}$), metallicity ([M/H]) and rotational velocity ($v\sin{i}$), using the Stellar Label Machine (SLAM; \citealt{2021GYJslam}), a data-driven forward spectral model based on support vector regression (SVR) \citep{2020zhangbolaspecslam}. 
It utilizes a training dataset constructed from TLUSTY synthetic spectra derived based on non-local thermodynamic equilibrium (NLTE) assumption \citep{2020zhangbolaspecslam}.
The low-resolution spectra from LAMOST offered broad wavelength coverage, enabling better constraints on $T_\mathrm{eff}$, $\log{g}$, and [M/H]. 
On the other hand, medium-resolution spectra (MRS) provided superior rotational velocity ($v\sin{i}$) estimations due to higher resolution and detailed line profiles \citep{2021GYJslam}.
Therefore, the values of $T_\mathrm{eff}$, $\log{g}$, and [M/H] were derived from LRS spectra, while $v\sin{i}$ was calculated from MRS spectra. 
The typical uncertainties for LAMOST MRS are $\sigma({T_{\rm eff}}) = 2,185$ K, $\sigma(V\sin{i}) = 11$ km s$^{-1}$ and $\sigma(\log{g}) = 0.29$ $\rm cm\ s^{-2}$, while for LAMOST LRS are $\sigma({T_{\rm eff}}) = 1,642$ K, $\sigma(V\sin{i}) = 42$ km s$^{-1}$ and $\sigma(\log{g}) = 0.25$ $\rm cm\ s^{-2}$ \citep{2021GYJslam}.
We omitted the broad diffuse interstellar bands of 4430, 5780, 6196, 6283, and 6614 \AA\ \citep{1995Herbig,2021GYJslam}.
We provide two examples from our sample, illustrating the observed spectra (black lines) along with the corresponding best-fitting spectra (red dashed lines) derived using SLAM. 

However, because the NLTE TLUSTY grid has a lower temperature bound of 15,000 K, using SLAM below this limit would involve extrapolation, which may introduce biases.
In contrast, \citet{2022Xiang} provide LTE atmospheric parameters based on ATLAS12 over 7,500--60,000 K, so the 9,000--15,000 K regime of our sample lies within their trained range, making those estimates more reliable (i.e., interpolation rather than extrapolation).
Although NLTE effects are significant for early-type stars \citep{2007LanztlustyB}, an analysis of LAMOST LRS spectra by \citet{2022Xiang} shows that, for temperatures below 25,000 K, LTE and NLTE determinations are in good agreement.
Accordingly, for stars in our sample with \( T_\mathrm{eff} \geq 15,000 \, \mathrm{K} \), atmospheric parameters were derived using SLAM, while for those with \( T_\mathrm{eff} < 15,000 \, \mathrm{K} \), we adopted the parameters provided by \cite{2022Xiang}.
The derived atmospheric parameters are listed in Table~\ref{Tab:1}.

\begin{figure}
    \centering
    \includegraphics[scale=0.6]{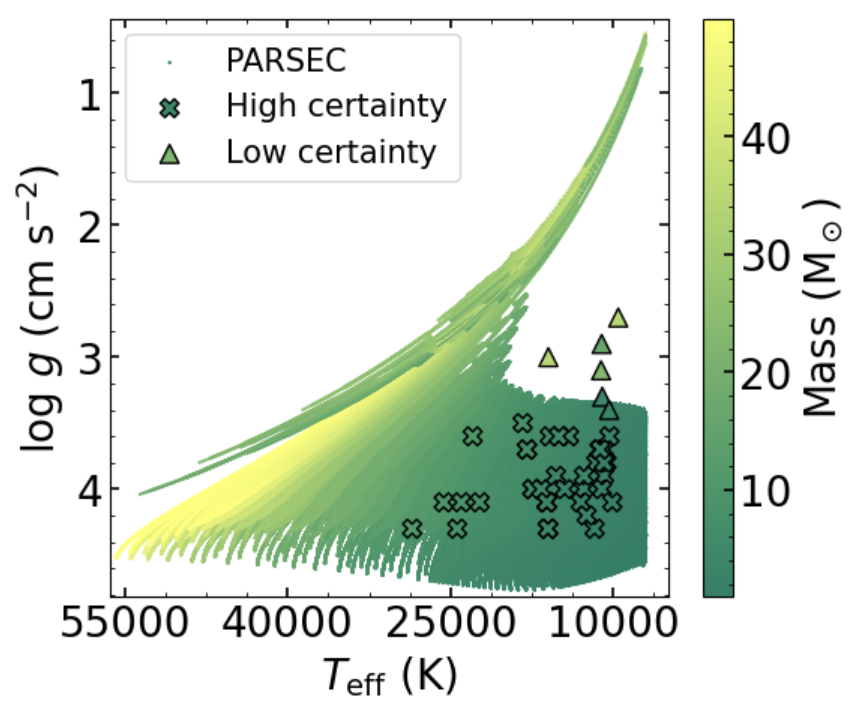}
    \caption{\G{Training grid and predicted sample in the $\log g$–$T_\mathrm{eff}$ plane.
The background shows the PARSEC~1.2S main-sequence grid color-coded by stellar mass.
The black crosses and triangles represent our 45 B-type runaway stars, with colors corresponding to their predicted masses using the same color scale.
Crosses denote stars lying within the parameter range of the training grid (high certainty), while triangles indicate extrapolated sources outside the grid (low certainty).}}
    \label{fig: MScolored}
\end{figure}

\subsection{\G{Physical parameters}}\label{sec: Fundamental parameters}
Using the atmospheric parameters obtained above, we derived the fundamental parameters, including masses and ages, for our sample following the method described by \citet{2025Liqida}.
This method employs a machine learning model implemented as a random-forest (RF) regression algorithm, which iteratively predicts stellar luminosity, mass and age based on atmospheric parameters. 
\G{RF regression is an ensemble method that pools the outputs of thousands of decision trees, each grown on a random subset of the training sample, to yield a single, robust prediction. When the atmospheric parameters are fed in, every decision tree makes its own prediction, and the RF then averages them to obtain the final value. This averaging dampens the idiosyncrasies of single trees, cutting variance and yielding stable, accurate results even when the underlying relations are noisy or highly non-linear \citep{2025Liqida}.}

\G{We used main-sequence stars extracted from the PARSEC 1.2S \citep{2012Bressan,2014Chen,2018Fu} as the training grid for the RF algorithm. The training grid comprises a grid of stellar parameters covering $7000\mathrm{K} < T_\mathrm{eff} < 49{,}000~\mathrm{K}$, $0.55 < \log g < 4.52$~dex, and $-0.6 < [\mathrm{M/H}] < 0.6$~dex. }
The process begins with the prediction of luminosity using $T_\mathrm{eff}$, $\log g$ and [M/H]. 
This predicted luminosity, along with $T_\mathrm{eff}$, $\log g$ and [M/H], is then used to predict stellar mass. 
Finally, the model predicts stellar age based on $T_\mathrm{eff}$, $\log g$, [M/H], as well as the previously derived luminosity and mass. 
\G{Validation of the method shows that the uncertainty of mass prediction is 9\%, whereas the age dispersion is 0.44 Gyr. The larger dispersion in age prediction arises from poorly constrained metallicity in early-type stars and the inherent age–metallicity degeneracy.}

The derived masses and ages are presented in Tab~\ref{Tab:1}.
\G{Figure~\ref{fig: MScolored} illustrates the PARSEC1.2S training grid in the $\log g$–$T_\mathrm{eff}$ plane, color-coded by stellar mass.
Our predicted sources are overplotted as black symbols using the same mass color scale.
Crosses mark stars located within the training grid (“high certainty”), whereas triangles correspond to extrapolated objects outside the grid (“low certainty”).
These two cases are also flagged as “I” (inside) and “O” (outside) in Table\ref{Tab:1}.
From the figure, six stars are found to lie outside the training grid — one of them have $T_\mathrm{eff}$ and $\log g$ within the grid boundaries but fall outside in [M/H].
\G{Since the RF model extrapolates for these out-of-range inputs, their predicted masses and ages are deemed unreliable, and sources marked with a triangle are also considered to lie outside the main-sequence phase.}
Four of these outliers exhibit unusually high inferred masses ($>16 M_\odot$), further supporting their exclusion.
Therefore, we removed these six stars from the subsequent analysis, resulting in a final sample of 39 B-type runaway stars.}

\subsection{Orbital trajectories}
Having derived the atmospheric and fundamental parameters, we use the Python package GALPY \citep{2015Bovygalpy} to calculate the Galactic trajectories of our stars. 
Proper motions were retrieved from Gaia EDR3 \citep{2021GaiaDr3}, and distances were obtained from the posterior distribution of geometric distances by \cite{2021Gaiadis}, while radial velocities and peculiar space velocities ($V_{\rm Sp}$) were taken from \cite{2024ApJSGYJ}. 
We adopt the solar motion of ($U_\sun$,$V_\sun$,$W_\sun$) = (11.10, 12.24, 7.25) km $\rm s^{-1}$ from \citet{2010Schonrich}, the solar Galactocentric distance $R_0$ = 8.5\,\text{kpc}, and the circular Galactic rotational velocity $V_c = 220$ km s$^{-1}$ from \citet{1986Kerr} (see \cite{2024ApJSGYJ}).
We determined the uncertainties in the birthplaces of the \G{39} stars by conducting Monte Carlo (MC) simulations, assuming Gaussian error distributions for distances, proper motions, radial velocities, and ages  \citep{2023liuzhicunB}. 

\begin{figure*}
    \centering
    \includegraphics[scale=0.5]{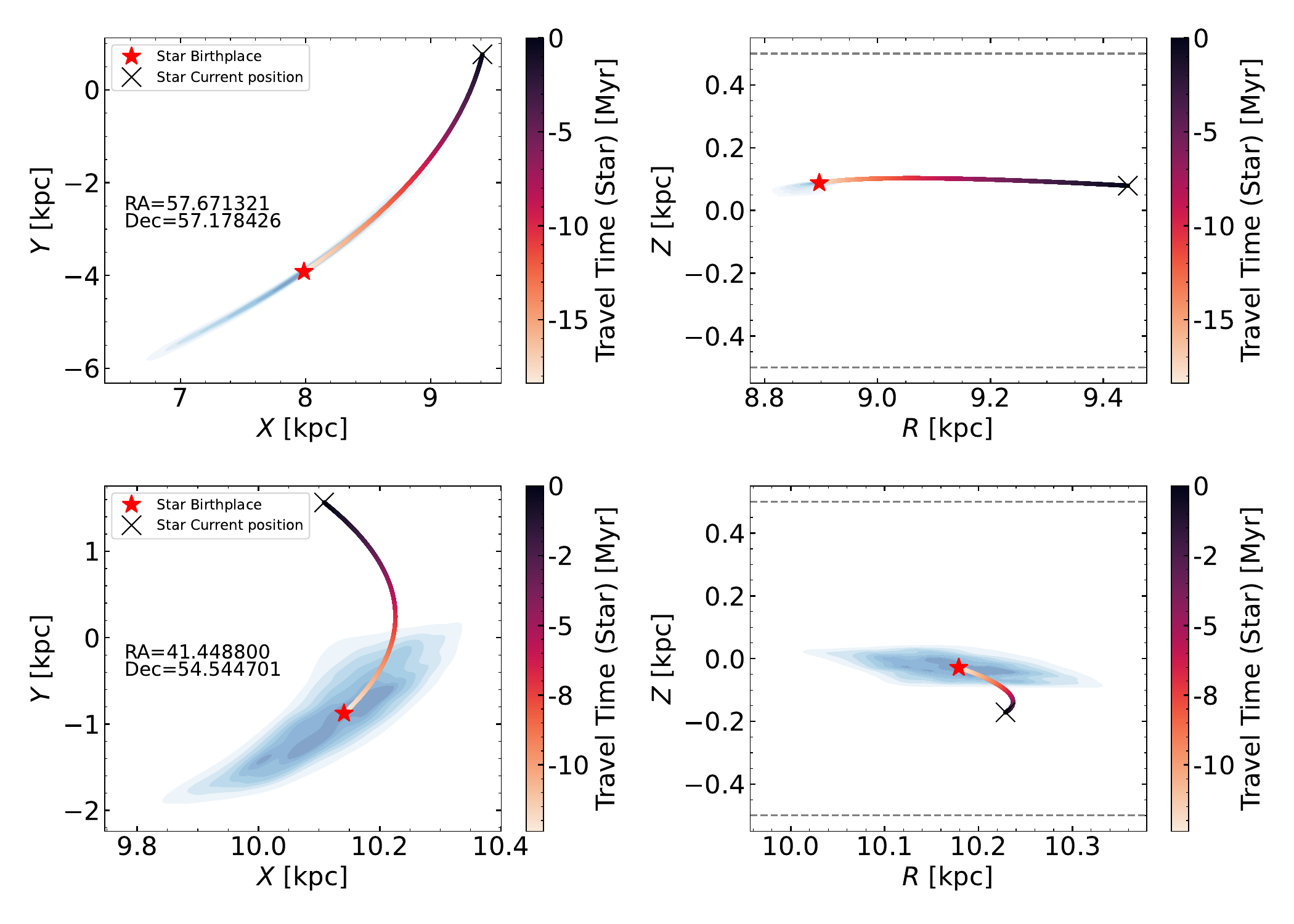}
    \caption{Orbital trajectories for two example stars with coordinates RA: 57.671321, Dec: 57.178426 (top panels) and RA: 41.448800, Dec: 54.544701 (bottom panels), shown in the Galactic $X-Y$ plane (left panels) and the Galactic $R-Z$ plane (right panels). The color scale indicates travel time (Myr), with the red star marking each star's birthplace and the black “x” marking its current position. The red-shaded contours denote the uncertainty regions of the birthplaces.}
    \label{fig: 32orbit}
\end{figure*}

\begin{figure*}
    \centering
    \includegraphics[scale=0.5]{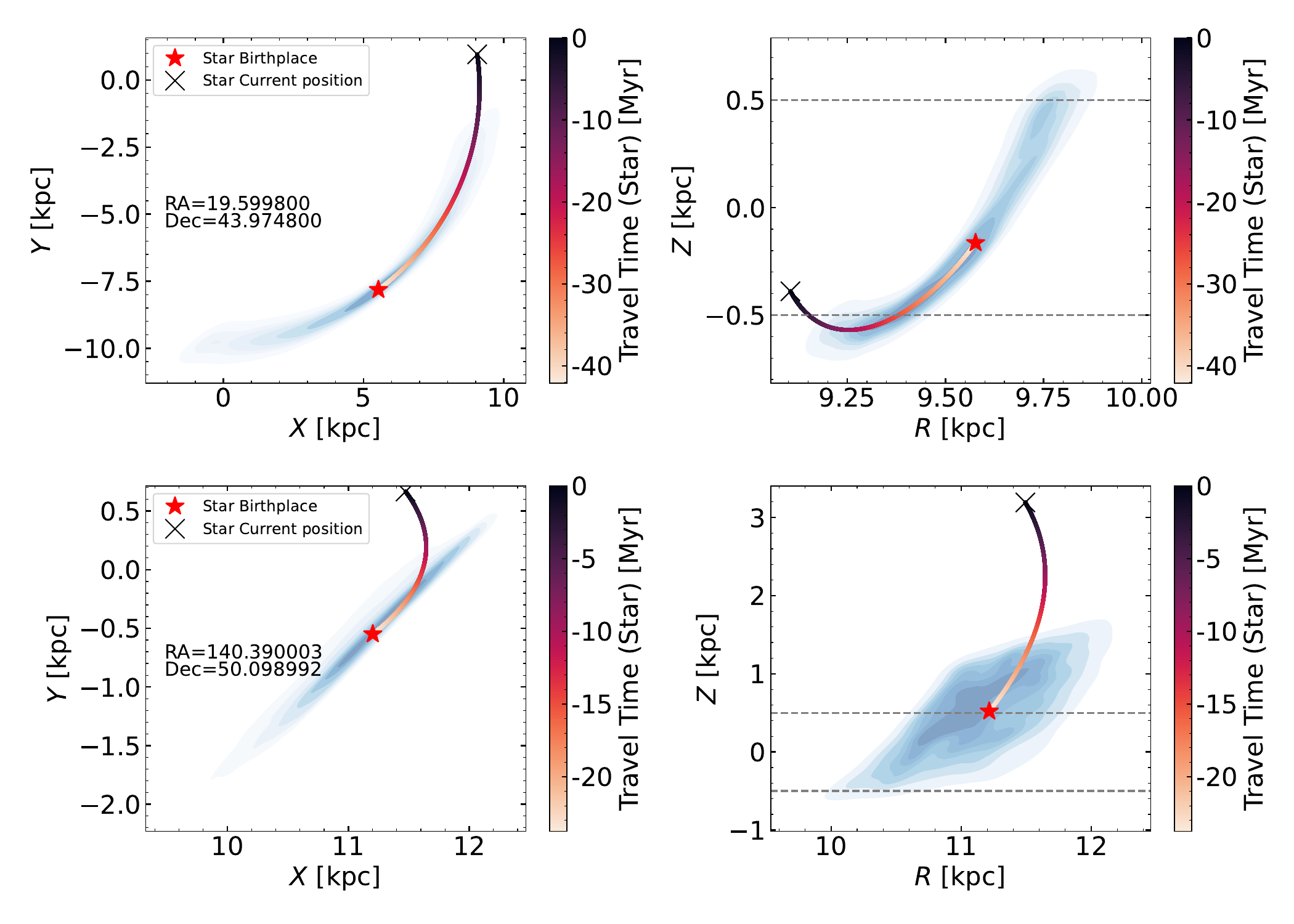}
    \caption{Orbital trajectories for two example stars with coordinates RA: 19.599800, Dec: 43.974800 (top panels) and RA: 140.390003, Dec: 50.098992 (bottom panels), shown in the Galactic $X-Y$ plane (left panels) and the Galactic $R-Z$ plane (right panels). The color scale indicates travel time (Myr), with the red star marking each star's birthplace and the black “x” marking its current position. The red-shaded contours denote the uncertainty regions of the inferred birthplaces.}
    \label{fig: 12orbit}
\end{figure*}

\section{Results} \label{sec:Results}
\subsection{Orbital trajectories of stars}

Although the Milky Way’s disk flare and warp cause the exponential scaleheight to increase with Galactocentric radius, reported values vary among studies because they use different populations and methods \citep{2014Corredoira,2018wanghaifeng,2021Yuyang}.
Based on LAMOST DR5 OB stars, \cite{2021Yuyang}  derived a scaleheight increasing from 0.14 to 0.5 kpc over $R$=8--14 kpc, similar to the thin disk values reported by \cite{2018wanghaifeng}.
For the thick disk, \cite{2014Corredoira} and \cite{2018wanghaifeng2} reported scaleheights of 0.5–1.5 kpc over $R$=8--20 kpc.
Considering the different scale heights of the thin and thick disks, and the R distribution of our sample, we adopt |$Z$|=0.5 kpc\footnote{\G{We also tested a smaller boundary of $|Z| = 0.14$~kpc and found that, when considering the uncertainties in the birth positions, all of our stars still originate from within the Galactic disk.}} as the threshold to distinguish disk objects \citep{2023liuzhicunB}.
\G{Notably, there is no rigid border orthogonal to the Galactic disk, and in a disk population with a given scale height, about 30\% of the stars may extend beyond that height.}

After computing the Galactic orbits of our stars, we present the trajectories for four example stars in Figures~\ref{fig: 32orbit} and \ref{fig: 12orbit}, shown in the Galactic $X-Y$ and $R-Z$ planes, with the color bar indicating travel time (Myr), defined here as the elapsed time for a star to move from its birthplace to its current position.
Red markers denote the birthplaces of the stars, black crosses indicate their current positions, and the red-shaded contours outline the uncertainties in the birth locations derived from Monte Carlo simulations.
Fig.~\ref{fig: 32orbit} shows two examples of stars for which there is no significant change between their current locations and their birthplaces and their orbits remain confined within $|Z| <$ 0.5 kpc, supporting a disk origin. 
In total, \G{29} stars exhibit this behavior.
In the top panel of Fig.~\ref{fig: 12orbit}, although the trajectories \G{pass through} the disk, the birthplaces of stars can still be traced back to within $|Z| <$ 0.5 kpc, and \G{9} stars in our sample fall into this category.
\G{Notably, we use the term ``pass through the disk'' (also referred to as ``disk-passing'' hereafter) to describe stars whose orbits pass the Galactic disk region \GG{(as defined here as being within |Z|<=0.5 kpc)}, which may involve passing through and leaving the disk, or possibly returning to it at a later time.}
In the bottom panel of Fig.~\ref{fig: 12orbit}, the birthplace of this star is closer to the disk edge, but after considering the MC uncertainties, they can still be regarded as having a disk origin.
We have \G{one} such sources in total.
In summary, \G{29} stars have trajectories entirely confined within the disk, while \G{10} stars have trajectories that \G{pass through} the disk but may still be considered to have a disk origin.

\subsection{Possible Cluster Origin}
\begin{figure*}
    \centering
    \includegraphics[scale=0.5]{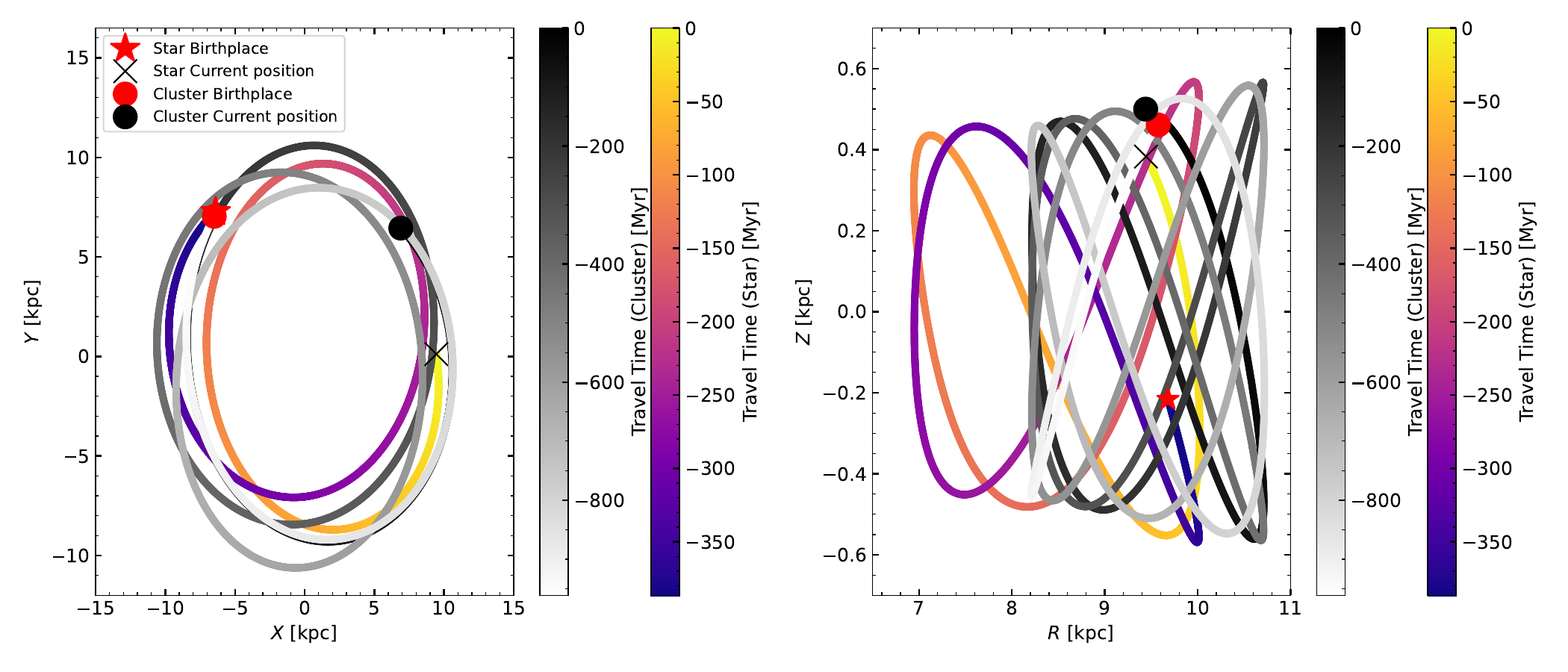}
    \caption{Orbital trajectories of IC 1311 and the star (RA = 102.263087, Dec = 41.981471) in Galactic coordinates.
The left and right panels show the orbits of the star (colored by travel time in Myr) and the cluster (grayscale) in the Galactic $X-Y$ and Galactic $R-Z$ projections, respectively. The red star and red circle mark the birthplaces of the star and cluster, while the black cross and black circle indicate their current positions. The colorbars indicate the look-back time along each orbit, with negative values corresponding to earlier times.}
    \label{fig: cluster1}
\end{figure*}

\begin{figure*}
    \centering
    \includegraphics[scale=0.5]{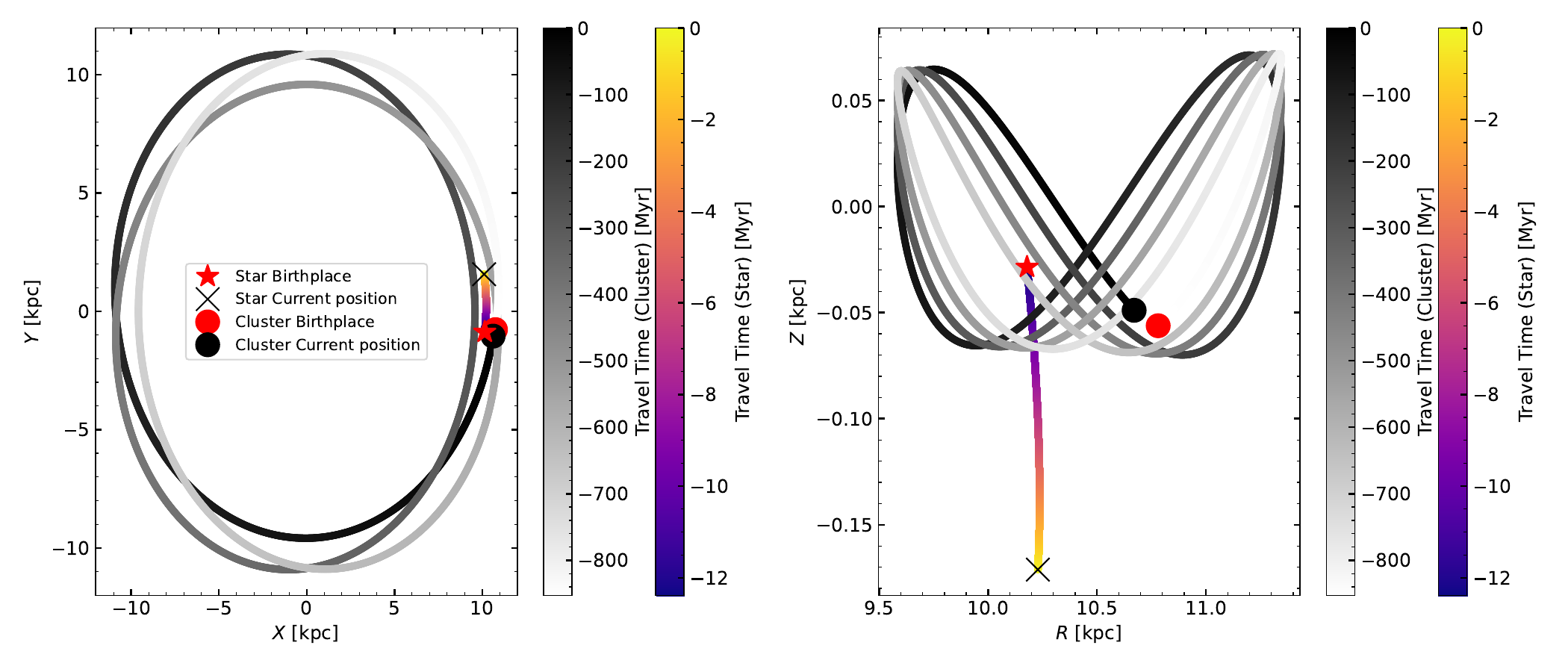}
    \caption{Orbital trajectories of NGC 2236 and the star (RA = 41.4488, Dec = 54.5447) in Galactic coordinates.
The left and right panels show the orbits of the star (colored by travel time in Myr) and the cluster (grayscale) in the Galactic $X-Y$ and Galactic $R-Z$ projections, respectively. The red star and red circle mark the birthplaces of the star  and cluster, while the black cross and black circle indicate their current positions. The colorbars indicate the look-back time along each orbit, with negative values corresponding to earlier times.}
    \label{fig: cluster2}
\end{figure*}

To investigate potential dynamical ejection scenarios (DES), we cross-matched our sample of \G{39} stars with cluster catalogs \citep{2019Castro,2020Castro,2023Hunt} and found that the trajectories of only two sources intersect with those of clusters.
We also use the Python package GALPY \citep{2015Bovygalpy} to calculate the Galactic trajectories for the two matched clusters and show the resluts in Fig.~\ref{fig: cluster1} and Fig.~\ref{fig: cluster2}.

Fig.~\ref{fig: cluster1} displays the orbital trajectories of a star with RA = 102.2630, Dec = 41.9814 and its matched cluster IC 1311. 
The left panel shows the Galactic $X$--$Y$ projection and the right panel shows the Galactic $R$--$Z$ projection in Galactic coordinates. 
The red star and red circle indicate the birthplaces of the star and cluster, while the black cross and black circle mark their present-day positions. 
Colorbars show the look-back time along each orbit.
The star has $T_\mathrm{eff}$ = 10,339 K, $\log g$ = 3.65, and [M/H] = –0.48, while the metallicity of IC 1311 is [M/H] = –0.30 $\pm$ 0.16 as reported by \citet{2009Warren}.
In the Galactic $X-Y$ plane, both the star and the cluster move counterclockwise, while in the $R-Z$ plane their trajectories broadly overlap.
The similarity between their orbital trajectories and metallicities suggests that the star may have originated from the cluster.

Fig.\ref{fig: cluster2} displays the orbital trajectories of a star with RA = 41.4488, Dec = 54.5447 and its matched cluster NGC 2236. 
The left panel shows the Galactic $X$--$Y$ projection and the right panel shows the Galactic $R$--$Z$ projection in Galactic coordinates, respectively. 
The marking scheme is the same as in Fig.\ref{fig: cluster1}.
The star has $T_\mathrm{eff}$ = 9,890 K, $\log g$ = 3.87, and [M/H] = –0.28. The metallicity of NGC 2236 is [M/H] = –0.19 $\pm$ 0.08 as reported by \citet{2020Zhong}. 
The orbital trajectories in Fig.~\ref{fig: cluster2} show that the star was initially close to NGC 2236, but later moved away from the cluster's orbit in the $R$--$Z$ projection.
This suggests that the star may have been ejected from the cluster.
Notably, definitive confirmation would require further evidence. 
As noted by \cite{2001Hoogerwerf}, compelling evidence for the dynamical ejection scenario (DES) involves finding a common site of origin for the individual components of the encounter.

\subsection{Statistical properties}

\begin{figure}
    \centering
    \includegraphics[scale=0.5]{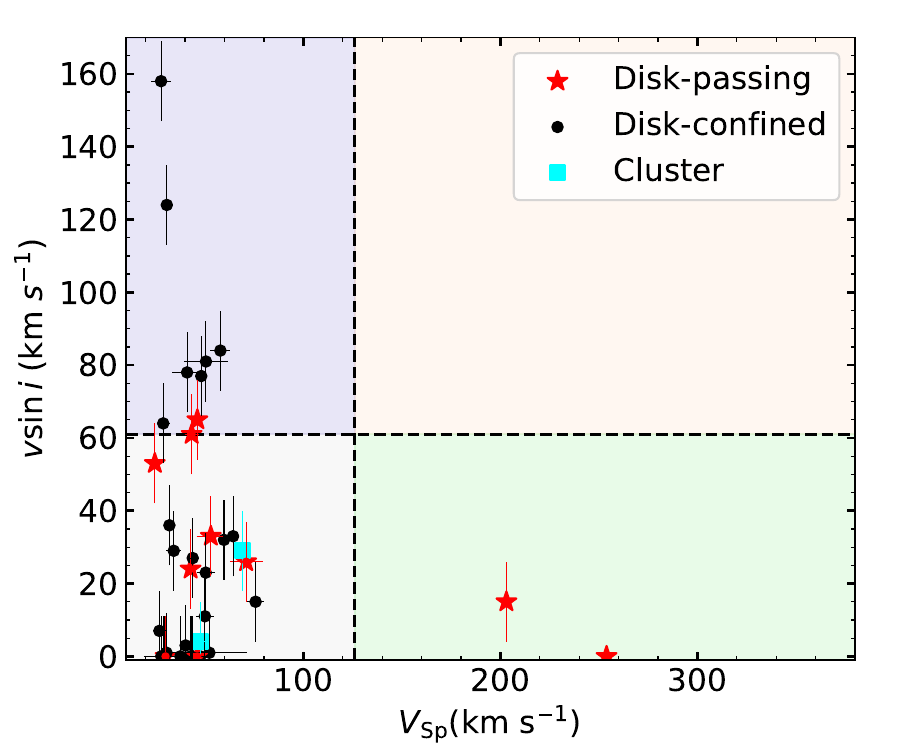}
    \caption{Distribution of runaway stars in the $V_{\rm Sp} - v\sin{i}$ plane.
Stars with orbits remaining within the Galactic disk are marked as black dots, those with \G{disk-passing} orbits are shown as red stars, and possible cluster-origin runaways are depicted as cyan squares.
Thresholds of $v\sin{i} = 61$ km s$^{-1}$ and $V_{\rm Sp} = 126$ km s$^{-1}$ (dashed lines) represent the $3\sigma$ values obtained from Maxwellian fits to the observed distributions, as described in Sec. 4.3 of \cite{2024ApJSGYJ}.}
    \label{fig: desert}
\end{figure}
Figure~\ref{fig: desert} presents the distribution of runaway stars in the $V_{\rm Sp} - v\sin{i}$ plane.
Black dots denote disk-confined runaways (orbits confined to the Galactic disk), whereas red stars mark disk-\G{passing} runaways.
Cyan squares denote runaway stars that may share a common origin with the cluster.
The dashed lines at $v\sin{i} = 61$ km s$^{-1}$ and $V_{\rm Sp} = 126$ km s$^{-1}$ correspond to the $3\sigma$ thresholds derived from Maxwellian fits to the respective distributions (Sec. 4.3 of \cite{2024ApJSGYJ}).

Figure~\ref{fig: desert} indicates that the majority of B-type runaways lie in the gray region, corresponding to low peculiar space velocities and low $v\sin{i}$.
The disk-confined subsample (black dots) has relatively low peculiar space velocities and spans a wide range of $v\sin{i}$, whereas the disk-\G{passing} subsample (red stars) tends to have higher $V_{\rm Sp}$ while maintaining low $v\sin{i}$.
According to \cite{2022Sana}, slow-moving, rapidly rotating runaways point to BSS, while fast-moving, slowly rotating runaways are more consistent with DES.
Rapidly rotating runaway stars are interpreted as products of close binary interactions, such as stellar mergers \citep{1992Podsiadlowski,2011Tylenda} or mass transfer \citep{1981Packet,2005Petrovic}, which can significantly increase the rotational velocity \citep{2022Sana}.
By contrast, DES involves close multi-body encounters that impart large $V_{\rm Sp}$ without significantly spinning up the ejected stars \citep{2022Sana}.
In addition, stars with a possible BSS origin tend to be disk-confined, whereas those with a possible DES origin are disk-\G{passing}.
While our orbital traceback provides statistical support for these trends, more definitive conclusions will require high-resolution spectroscopic data in future work.

\section{Summary} \label{sec:Summary}
Runaway stars rapidly escape from their birth clusters or associations, typically with peculiar velocities of 30--40~km/s. 
Two mechanisms are typically considered, the binary supernova scenario and the dynamical ejection scenario, but their relative contributions remain uncertain.
Determining the origins of runaways may provide constraints on binary evolution and dynamical processes.

In this work, we investigate the kinematic origins of \G{39} B--type runaway stars identified by \citet{2024ApJSGYJ}. 
We determined spectral subclasses from key absorption lines and derived atmospheric parameters with SLAM. We then estimated fundamental parameters using a machine learning model trained on these parameters.
Using GALPY, we compute each star's Galactic trajectory and find that \G{29} stars remain fully confined to the disk, while \G{10 have disk-passing} orbits yet still trace back to it.
Two of these may be consistent with a cluster origin, although definitive confirmation would require further evidence.
We also find  a feature in the $V_{\rm Sp} - v\sin{i}$ plane: low-$V_{\rm Sp}$, high-$v\sin{i}$ runaways are typically disk-confined, whereas high-$V_{\rm Sp}$, low-$v\sin{i}$ runaways are disk-\G{passing}, probably reflecting two different origins.
This may provide new kinematic evidence to distinguish between the binary supernova and dynamical ejection scenarios.
However, our conclusions are limited by the lack of high-resolution spectroscopic abundances. Future surveys with higher precision spectroscopy will be crucial to establish firmer constraints on the origins of runaway stars and their role in binary evolution and cluster dynamics.

\section{Data availability}
Table 1 is only available in electronic form at the CDS via anonymous ftp to cdsarc.u-strasbg.fr (130.79.128.5) or via http://cdsweb.u-strasbg.fr/cgi-bin/qcat?J/A+A/.

\begin{acknowledgements}
This work is supported by the National Natural Science Foundation of China (Nos.\ 12288102, 12125303, 12090040/3, 12103064, 12403039, 12373036), the National Key R\&D Program of China (grant Nos. 2021YFA1600403/1, 2021YFA1600400), and the Natural Science Foundation of Yunnan Province (Nos. 202201BC070003, 202001AW070007), the International Centre of Supernovae, Yunnan Key Laboratory (No. 202302AN360001), the “Yunnan Revitalization Talent Support Program"-Science, Technology Champion Project (N0. 202305AB350003), and Yunnan Fundamental Research Projects (grant Nos. 202501CF070018).
\end{acknowledgements}

\bibliographystyle{aa} 
\bibliography{aa57420-25}

@ARTICLE{2020zhangbolaspecslam,
       author = {{Zhang}, Bo and {Liu}, Chao and {Deng}, Li-Cai},
        title = "{Deriving the Stellar Labels of LAMOST Spectra with the Stellar LAbel Machine (SLAM)}",
      journal = {\apjs},
         year = 2020,
        month = jan,
       volume = {246},
       number = {1},
          eid = {9},
        pages = {9},
          doi = {10.3847/1538-4365/ab55ef},
archivePrefix = {arXiv},
       eprint = {1908.08677},
 primaryClass = {astro-ph.SR},
       adsurl = {https://ui.adsabs.harvard.edu/abs/2020ApJS..246....9Z}
}

@ARTICLE{2012CuiXiangQun,
       author = {{Cui}, Xiang-Qun and {Zhao}, Yong-Heng and {Chu}, Yao-Quan and
         {Li}, Guo-Ping and {Li}, Qi and {Zhang}, Li-Ping and {Su}, Hong-Jun and
         {Yao}, Zheng-Qiu and {Wang}, Ya-Nan and {Xing}, Xiao-Zheng and
         {Li}, Xin-Nan and {Zhu}, Yong-Tian and {Wang}, Gang and {Gu}, Bo-Zhong and
         {Luo}, A. -Li and {Xu}, Xin-Qi and {Zhang}, Zhen-Chao and
         {Liu}, Gen-Rong and {Zhang}, Hao-Tong and {Yang}, De-Hua and
         {Cao}, Shu-Yun and {Chen}, Hai-Yuan and {Chen}, Jian-Jun and
         {Chen}, Kun-Xin and {Chen}, Ying and {Chu}, Jia-Ru and {Feng}, Lei and
         {Gong}, Xue-Fei and {Hou}, Yong-Hui and {Hu}, Hong-Zhuan and
         {Hu}, Ning-Sheng and {Hu}, Zhong-Wen and {Jia}, Lei and
         {Jiang}, Fang-Hua and {Jiang}, Xiang and {Jiang}, Zi-Bo and {Jin}, Ge and
         {Li}, Ai-Hua and {Li}, Yan and {Li}, Ye-Ping and {Liu}, Guan-Qun and
         {Liu}, Zhi-Gang and {Lu}, Wen-Zhi and {Mao}, Yin-Dun and {Men}, Li and
         {Qi}, Yong-Jun and {Qi}, Zhao-Xiang and {Shi}, Huo-Ming and
         {Tang}, Zheng-Hong and {Tao}, Qing-Sheng and {Wang}, Da-Qi and
         {Wang}, Dan and {Wang}, Guo-Min and {Wang}, Hai and {Wang}, Jia-Ning and
         {Wang}, Jian and {Wang}, Jian-Ling and {Wang}, Jian-Ping and
         {Wang}, Lei and {Wang}, Shu-Qing and {Wang}, You and {Wang}, Yue-Fei and
         {Xu}, Ling-Zhe and {Xu}, Yan and {Yang}, Shi-Hai and {Yu}, Yong and
         {Yuan}, Hui and {Yuan}, Xiang-Yan and {Zhai}, Chao and {Zhang}, Jing and
         {Zhang}, Yan-Xia and {Zhang}, Yong and {Zhao}, Ming and {Zhou}, Fang and
         {Zhou}, Guo-Hua and {Zhu}, Jie and {Zou}, Si-Cheng},
        title = "{The Large Sky Area Multi-Object Fiber Spectroscopic Telescope (LAMOST)}",
      journal = {Research in Astronomy and Astrophysics},
         year = 2012,
        month = sep,
       volume = {12},
       number = {9},
        pages = {1197-1242},
          doi = {10.1088/1674-4527/12/9/003},
       adsurl = {https://ui.adsabs.harvard.edu/abs/2012RAA....12.1197C}
}

@ARTICLE{2012ZhaoGang,
       author = {{Zhao}, Gang and {Zhao}, Yong-Heng and {Chu}, Yao-Quan and
         {Jing}, Yi-Peng and {Deng}, Li-Cai},
        title = "{LAMOST spectral survey {\textemdash} An overview}",
      journal = {Research in Astronomy and Astrophysics},
         year = 2012,
        month = jul,
       volume = {12},
       number = {7},
        pages = {723-734},
          doi = {10.1088/1674-4527/12/7/002},
       adsurl = {https://ui.adsabs.harvard.edu/abs/2012RAA....12..723Z}
}

@ARTICLE{2012DengLiCai,
       author = {{Deng}, Li-Cai and {Newberg}, Heidi Jo and {Liu}, Chao and
         {Carlin}, Jeffrey L. and {Beers}, Timothy C. and {Chen}, Li and
         {Chen}, Yu-Qin and {Christlieb}, Norbert and {Grillmair}, Carl J. and
         {Guhathakurta}, Puragra and {Han}, Zhan-Wen and {Hou}, Jin-Liang and
         {Lee}, Hsu-Tai and {L{\'e}pine}, S{\'e}bastien and {Li}, Jing and
         {Liu}, Xiao-Wei and {Pan}, Kai-Ke and {Sellwood}, J.~A. and {Wang}, Bo and
         {Wang}, Hong-Chi and {Yang}, Fan and {Yanny}, Brian and
         {Zhang}, Hao-Tong and {Zhang}, Yue-Yang and {Zheng}, Zheng and
         {Zhu}, Zi},
        title = "{LAMOST Experiment for Galactic Understanding and Exploration (LEGUE) {\textemdash} The survey's science plan}",
      journal = {Research in Astronomy and Astrophysics},
         year = 2012,
        month = jul,
       volume = {12},
       number = {7},
        pages = {735-754},
          doi = {10.1088/1674-4527/12/7/003},
archivePrefix = {arXiv},
       eprint = {1206.3578},
 primaryClass = {astro-ph.GA},
       adsurl = {https://ui.adsabs.harvard.edu/abs/2012RAA....12..735D}
}

@ARTICLE{2020LiuChao,
       author = {{Liu}, Chao and {Fu}, Jianning and {Shi}, Jianrong and {Wu}, Hong and
         {Han}, Zhanwen and {Chen}, Li and {Dong}, Subo and {Zhao}, Yongheng and
         {Chen}, Jian-Jun and {Zhang}, Haotong and {Bai}, Zhong-Rui and
         {Chen}, Xuefei and {Cui}, Wenyuan and {Du}, Bing and {Hsia}, Chih-Hao and
         {Jiang}, Deng-Kai and {Hou}, Jinliang and {Hou}, Wen and {Li}, Haining and
         {Li}, Jiao and {Li}, Lifang and {Liu}, Jiaming and {Liu}, Jifeng and
         {Luo}, A-Li and {Ren}, Juan-Juan and {Tian}, Hai-Jun and {Tian}, Hao and
         {Wang}, Jia-Xin and {Wu}, Chao-Jian and {Xie}, Ji-Wei and
         {Yan}, Hong-Liang and {Yang}, Fan and {Yu}, Jincheng and {Zhang}, Bo and
         {Zhang}, Huawei and {Zhang}, Li-Yun and {Zhang}, Wei and {Zhao}, Gang and
         {Zhong}, Jing and {Zong}, Weikai and {Zuo}, Fang},
        title = "{LAMOST Medium-Resolution Spectroscopic Survey (LAMOST-MRS): Scientific goals and survey plan}",
      journal = {arXiv e-prints},
         year = 2020,
        month = may,
          eid = {arXiv:2005.07210},
        pages = {arXiv:2005.07210},
archivePrefix = {arXiv},
       eprint = {2005.07210},
 primaryClass = {astro-ph.SR},
       adsurl = {https://ui.adsabs.harvard.edu/abs/2020arXiv200507210L}
}

@ARTICLE{2023gaianewrunaway,
       author = {{Carretero-Castrillo}, M. and {Rib{\'o}}, M. and {Paredes}, J.~M.},
        title = "{Galactic runaway O and Be stars found using Gaia DR3}",
      journal = {\aap},
         year = 2023,
        month = nov,
       volume = {679},
          eid = {A109},
        pages = {A109},
          doi = {10.1051/0004-6361/202346613},
archivePrefix = {arXiv},
       eprint = {2311.01827},
 primaryClass = {astro-ph.SR},
       adsurl = {https://ui.adsabs.harvard.edu/abs/2023A&A...679A.109C}
}

@ARTICLE{2019LiuZhicun,
       author = {{Liu}, Zhicun and {Cui}, Wenyuan and {Liu}, Chao and {Huang}, Yang and
         {Zhao}, Gang and {Zhang}, Bo},
        title = "{A Catalog of OB Stars from LAMOST Spectroscopic Survey}",
      journal = {\apjs},
         year = 2019,
        month = apr,
       volume = {241},
       number = {2},
          eid = {32},
        pages = {32},
          doi = {10.3847/1538-4365/ab0a0d},
archivePrefix = {arXiv},
       eprint = {1902.07607},
 primaryClass = {astro-ph.SR},
       adsurl = {https://ui.adsabs.harvard.edu/abs/2019ApJS..241...32L}
}

@ARTICLE{2021GYJslam,
       author = {{Guo}, Yanjun and {Zhang}, Bo and {Liu}, Chao and {Li}, Jiao and {Li}, Jiangdan and {Wang}, Luqian and {Liu}, Zhicun and {Hou}, Yong-Hui and {Han}, Zhanwen and {Chen}, Xuefei},
        title = "{The Early-type Stars from the LAMOST Survey: Atmospheric Parameters}",
      journal = {\apjs},
         year = 2021,
        month = dec,
       volume = {257},
       number = {2},
          eid = {54},
        pages = {54},
          doi = {10.3847/1538-4365/ac2ded},
archivePrefix = {arXiv},
       eprint = {2110.06246},
 primaryClass = {astro-ph.SR},
       adsurl = {https://ui.adsabs.harvard.edu/abs/2021ApJS..257...54G}
}

@ARTICLE{2022luqianBe,
       author = {{Wang}, Luqian and {Li}, Jiao and {Wu}, You and {Gies}, Douglas R. and {Liu}, Jin Zhong and {Liu}, Chao and {Guo}, Yanjun and {Chen}, Xuefei and {Han}, Zhanwen},
        title = "{Identification of New Classical Be Stars from the LAMOST Medium Resolution Survey}",
      journal = {\apjs},
         year = 2022,
        month = jun,
       volume = {260},
       number = {2},
          eid = {35},
        pages = {35},
          doi = {10.3847/1538-4365/ac617a},
archivePrefix = {arXiv},
       eprint = {2203.15289},
 primaryClass = {astro-ph.SR},
       adsurl = {https://ui.adsabs.harvard.edu/abs/2022ApJS..260...35W}
}

@ARTICLE{1961Blaauw,
       author = {{Blaauw}, A.},
        title = "{On the origin of the O- and B-type stars with high velocities (the ``run-away'' stars), and some related problems}",
      journal = {\bain},
         year = 1961,
        month = may,
       volume = {15},
        pages = {265},
       adsurl = {https://ui.adsabs.harvard.edu/abs/1961BAN....15..265B}
}

@ARTICLE{1967Poveda,
       author = {{Poveda}, A. and {Ruiz}, J. and {Allen}, C.},
        title = "{Run-away Stars as the Result of the Gravitational Collapse of Proto-stellar Clusters}",
      journal = {Boletin de los Observatorios Tonantzintla y Tacubaya},
         year = 1967,
        month = apr,
       volume = {4},
        pages = {86-90},
       adsurl = {https://ui.adsabs.harvard.edu/abs/1967BOTT....4...86P}
}

@ARTICLE{2010Schonrich,
       author = {{Sch{\"o}nrich}, Ralph and {Binney}, James and {Dehnen}, Walter},
        title = "{Local kinematics and the local standard of rest}",
      journal = {\mnras},
         year = 2010,
        month = apr,
       volume = {403},
       number = {4},
        pages = {1829-1833},
          doi = {10.1111/j.1365-2966.2010.16253.x},
archivePrefix = {arXiv},
       eprint = {0912.3693},
 primaryClass = {astro-ph.GA},
       adsurl = {https://ui.adsabs.harvard.edu/abs/2010MNRAS.403.1829S}
}

@ARTICLE{1986Kerr,
       author = {{Kerr}, F.~J. and {Lynden-Bell}, D.},
        title = "{Review of galactic constants.}",
      journal = {\mnras},
         year = 1986,
        month = aug,
       volume = {221},
        pages = {1023-1038},
          doi = {10.1093/mnras/221.4.1023},
       adsurl = {https://ui.adsabs.harvard.edu/abs/1986MNRAS.221.1023K}
}

@ARTICLE{2005Mdzinarishvili,
       author = {{Mdzinarishvili}, T.~G. and {Chargeishvili}, K.~B.},
        title = "{New runaway OB stars with HIPPARCOS}",
      journal = {\aap},
         year = 2005,
        month = feb,
       volume = {431},
        pages = {L1-L4},
          doi = {10.1051/0004-6361:200400134},
       adsurl = {https://ui.adsabs.harvard.edu/abs/2005A&A...431L...1M}
}

@ARTICLE{Gies1986,
       author = {{Gies}, D.~R. and {Bolton}, C.~T.},
        title = "{The Binary Frequency and Origin of the OB Runaway Stars}",
      journal = {\apjs},
         year = 1986,
        month = jun,
       volume = {61},
        pages = {419},
          doi = {10.1086/191118},
       adsurl = {https://ui.adsabs.harvard.edu/abs/1986ApJS...61..419G}
}

@ARTICLE{Berger2001,
       author = {{Berger}, D.~H. and {Gies}, D.~R.},
        title = "{A Search for High-Velocity Be Stars}",
      journal = {\apj},
         year = 2001,
        month = jul,
       volume = {555},
       number = {1},
        pages = {364-367},
          doi = {10.1086/321461},
       adsurl = {https://ui.adsabs.harvard.edu/abs/2001ApJ...555..364B}
}

@ARTICLE{2022Sana,
       author = {{Sana}, H. and {Ram{\'\i}rez-Agudelo}, O.~H. and {H{\'e}nault-Brunet}, V. and {Mahy}, L. and {Almeida}, L.~A. and {de Koter}, A. and {Bestenlehner}, J.~M. and {Evans}, C.~J. and {Langer}, N. and {Schneider}, F.~R.~N. and {Crowther}, P.~A. and {de Mink}, S.~E. and {Herrero}, A. and {Lennon}, D.~J. and {Gieles}, M. and {Ma{\'\i}z Apell{\'a}niz}, J. and {Renzo}, M. and {Sabbi}, E. and {van Loon}, J. Th. and {Vink}, J.~S.},
        title = "{The VLT-FLAMES Tarantula Survey. Observational evidence for two distinct populations of massive runaway stars in 30 Doradus}",
      journal = {\aap},
         year = 2022,
        month = dec,
       volume = {668},
          eid = {L5},
        pages = {L5},
          doi = {10.1051/0004-6361/202244677},
archivePrefix = {arXiv},
       eprint = {2211.13476},
 primaryClass = {astro-ph.SR},
       adsurl = {https://ui.adsabs.harvard.edu/abs/2022A&A...668L...5S}
}

@ARTICLE{2021GaiaDr3,
       author = {{Gaia Collaboration} and {Brown}, A.~G.~A. and {Vallenari}, A. and {Prusti}, T. and {de Bruijne}, J.~H.~J. and {Babusiaux}, C. and {Biermann}, M. and {Creevey}, O.~L. and {Evans}, D.~W. and {Eyer}, L. and {Hutton}, A. and {Jansen}, F. and {Jordi}, C. and {Klioner}, S.~A. and {Lammers}, U. and {Lindegren}, L. and {Luri}, X. and {Mignard}, F. and {Panem}, C. and {Pourbaix}, D. and {Randich}, S. and {Sartoretti}, P. and {Soubiran}, C. and {Walton}, N.~A. and {Arenou}, F. and {Bailer-Jones}, C.~A.~L. and {Bastian}, U. and {Cropper}, M. and {Drimmel}, R. and {Katz}, D. and {Lattanzi}, M.~G. and {van Leeuwen}, F. and {Bakker}, J. and {Cacciari}, C. and {Casta{\~n}eda}, J. and {De Angeli}, F. and {Ducourant}, C. and {Fabricius}, C. and {Fouesneau}, M. and {Fr{\'e}mat}, Y. and {Guerra}, R. and {Guerrier}, A. and {Guiraud}, J. and {Jean-Antoine Piccolo}, A. and {Masana}, E. and {Messineo}, R. and {Mowlavi}, N. and {Nicolas}, C. and {Nienartowicz}, K. and {Pailler}, F. and {Panuzzo}, P. and {Riclet}, F. and {Roux}, W. and {Seabroke}, G.~M. and {Sordo}, R. and {Tanga}, P. and {Th{\'e}venin}, F. and {Gracia-Abril}, G. and {Portell}, J. and {Teyssier}, D. and {Altmann}, M. and {Andrae}, R. and {Bellas-Velidis}, I. and {Benson}, K. and {Berthier}, J. and {Blomme}, R. and {Brugaletta}, E. and {Burgess}, P.~W. and {Busso}, G. and {Carry}, B. and {Cellino}, A. and {Cheek}, N. and {Clementini}, G. and {Damerdji}, Y. and {Davidson}, M. and {Delchambre}, L. and {Dell'Oro}, A. and {Fern{\'a}ndez-Hern{\'a}ndez}, J. and {Galluccio}, L. and {Garc{\'\i}a-Lario}, P. and {Garcia-Reinaldos}, M. and {Gonz{\'a}lez-N{\'u}{\~n}ez}, J. and {Gosset}, E. and {Haigron}, R. and {Halbwachs}, J. -L. and {Hambly}, N.~C. and {Harrison}, D.~L. and {Hatzidimitriou}, D. and {Heiter}, U. and {Hern{\'a}ndez}, J. and {Hestroffer}, D. and {Hodgkin}, S.~T. and {Holl}, B. and {Jan{\ss}en}, K. and {Jevardat de Fombelle}, G. and {Jordan}, S. and {Krone-Martins}, A. and {Lanzafame}, A.~C. and {L{\"o}ffler}, W. and {Lorca}, A. and {Manteiga}, M. and {Marchal}, O. and {Marrese}, P.~M. and {Moitinho}, A. and {Mora}, A. and {Muinonen}, K. and {Osborne}, P. and {Pancino}, E. and {Pauwels}, T. and {Petit}, J. -M. and {Recio-Blanco}, A. and {Richards}, P.~J. and {Riello}, M. and {Rimoldini}, L. and {Robin}, A.~C. and {Roegiers}, T. and {Rybizki}, J. and {Sarro}, L.~M. and {Siopis}, C. and {Smith}, M. and {Sozzetti}, A. and {Ulla}, A. and {Utrilla}, E. and {van Leeuwen}, M. and {van Reeven}, W. and {Abbas}, U. and {Abreu Aramburu}, A. and {Accart}, S. and {Aerts}, C. and {Aguado}, J.~J. and {Ajaj}, M. and {Altavilla}, G. and {{\'A}lvarez}, M.~A. and {{\'A}lvarez Cid-Fuentes}, J. and {Alves}, J. and {Anderson}, R.~I. and {Anglada Varela}, E. and {Antoja}, T. and {Audard}, M. and {Baines}, D. and {Baker}, S.~G. and {Balaguer-N{\'u}{\~n}ez}, L. and {Balbinot}, E. and {Balog}, Z. and {Barache}, C. and {Barbato}, D. and {Barros}, M. and {Barstow}, M.~A. and {Bartolom{\'e}}, S. and {Bassilana}, J. -L. and {Bauchet}, N. and {Baudesson-Stella}, A. and {Becciani}, U. and {Bellazzini}, M. and {Bernet}, M. and {Bertone}, S. and {Bianchi}, L. and {Blanco-Cuaresma}, S. and {Boch}, T. and {Bombrun}, A. and {Bossini}, D. and {Bouquillon}, S. and {Bragaglia}, A. and {Bramante}, L. and {Breedt}, E. and {Bressan}, A. and {Brouillet}, N. and {Bucciarelli}, B. and {Burlacu}, A. and {Busonero}, D. and {Butkevich}, A.~G. and {Buzzi}, R. and {Caffau}, E. and {Cancelliere}, R. and {C{\'a}novas}, H. and {Cantat-Gaudin}, T. and {Carballo}, R. and {Carlucci}, T. and {Carnerero}, M.~I. and {Carrasco}, J.~M. and {Casamiquela}, L. and {Castellani}, M. and {Castro-Ginard}, A. and {Castro Sampol}, P. and {Chaoul}, L. and {Charlot}, P. and {Chemin}, L. and {Chiavassa}, A. and {Cioni}, M. -R.~L. and {Comoretto}, G. and {Cooper}, W.~J. and {Cornez}, T. and {Cowell}, S. and {Crifo}, F. and {Crosta}, M. and {Crowley}, C. and {Dafonte}, C. and {Dapergolas}, A. and {David}, M. and {David}, P. and {de Laverny}, P. and {De Luise}, F. and {De March}, R. and {De Ridder}, J. and {de Souza}, R. and {de Teodoro}, P. and {de Torres}, A. and {del Peloso}, E.~F. and {del Pozo}, E. and {Delbo}, M. and {Delgado}, A. and {Delgado}, H.~E. and {Delisle}, J. -B. and {Di Matteo}, P. and {Diakite}, S. and {Diener}, C. and {Distefano}, E. and {Dolding}, C. and {Eappachen}, D. and {Edvardsson}, B. and {Enke}, H. and {Esquej}, P. and {Fabre}, C. and {Fabrizio}, M. and {Faigler}, S. and {Fedorets}, G. and {Fernique}, P. and {Fienga}, A. and {Figueras}, F. and {Fouron}, C. and {Fragkoudi}, F. and {Fraile}, E. and {Franke}, F. and {Gai}, M. and {Garabato}, D. and {Garcia-Gutierrez}, A. and {Garc{\'\i}a-Torres}, M. and {Garofalo}, A. and {Gavras}, P. and {Gerlach}, E. and {Geyer}, R. and {Giacobbe}, P. and {Gilmore}, G. and {Girona}, S. and {Giuffrida}, G. and {Gomel}, R. and {Gomez}, A. and {Gonzalez-Santamaria}, I. and {Gonz{\'a}lez-Vidal}, J.~J. and {Granvik}, M. and {Guti{\'e}rrez-S{\'a}nchez}, R. and {Guy}, L.~P. and {Hauser}, M. and {Haywood}, M. and {Helmi}, A. and {Hidalgo}, S.~L. and {Hilger}, T. and {H{\l}adczuk}, N. and {Hobbs}, D. and {Holland}, G. and {Huckle}, H.~E. and {Jasniewicz}, G. and {Jonker}, P.~G. and {Juaristi Campillo}, J. and {Julbe}, F. and {Karbevska}, L. and {Kervella}, P. and {Khanna}, S. and {Kochoska}, A. and {Kontizas}, M. and {Kordopatis}, G. and {Korn}, A.~J. and {Kostrzewa-Rutkowska}, Z. and {Kruszy{\'n}ska}, K. and {Lambert}, S. and {Lanza}, A.~F. and {Lasne}, Y. and {Le Campion}, J. -F. and {Le Fustec}, Y. and {Lebreton}, Y. and {Lebzelter}, T. and {Leccia}, S. and {Leclerc}, N. and {Lecoeur-Taibi}, I. and {Liao}, S. and {Licata}, E. and {Lindstr{\o}m}, E.~P. and {Lister}, T.~A. and {Livanou}, E. and {Lobel}, A. and {Madrero Pardo}, P. and {Managau}, S. and {Mann}, R.~G. and {Marchant}, J.~M. and {Marconi}, M. and {Marcos Santos}, M.~M.~S. and {Marinoni}, S. and {Marocco}, F. and {Marshall}, D.~J. and {Martin Polo}, L. and {Mart{\'\i}n-Fleitas}, J.~M. and {Masip}, A. and {Massari}, D. and {Mastrobuono-Battisti}, A. and {Mazeh}, T. and {McMillan}, P.~J. and {Messina}, S. and {Michalik}, D. and {Millar}, N.~R. and {Mints}, A. and {Molina}, D. and {Molinaro}, R. and {Moln{\'a}r}, L. and {Montegriffo}, P. and {Mor}, R. and {Morbidelli}, R. and {Morel}, T. and {Morris}, D. and {Mulone}, A.~F. and {Munoz}, D. and {Muraveva}, T. and {Murphy}, C.~P. and {Musella}, I. and {Noval}, L. and {Ord{\'e}novic}, C. and {Orr{\`u}}, G. and {Osinde}, J. and {Pagani}, C. and {Pagano}, I. and {Palaversa}, L. and {Palicio}, P.~A. and {Panahi}, A. and {Pawlak}, M. and {Pe{\~n}alosa Esteller}, X. and {Penttil{\"a}}, A. and {Piersimoni}, A.~M. and {Pineau}, F. -X. and {Plachy}, E. and {Plum}, G. and {Poggio}, E. and {Poretti}, E. and {Poujoulet}, E. and {Pr{\v{s}}a}, A. and {Pulone}, L. and {Racero}, E. and {Ragaini}, S. and {Rainer}, M. and {Raiteri}, C.~M. and {Rambaux}, N. and {Ramos}, P. and {Ramos-Lerate}, M. and {Re Fiorentin}, P. and {Regibo}, S. and {Reyl{\'e}}, C. and {Ripepi}, V. and {Riva}, A. and {Rixon}, G. and {Robichon}, N. and {Robin}, C. and {Roelens}, M. and {Rohrbasser}, L. and {Romero-G{\'o}mez}, M. and {Rowell}, N. and {Royer}, F. and {Rybicki}, K.~A. and {Sadowski}, G. and {Sagrist{\`a} Sell{\'e}s}, A. and {Sahlmann}, J. and {Salgado}, J. and {Salguero}, E. and {Samaras}, N. and {Sanchez Gimenez}, V. and {Sanna}, N. and {Santove{\~n}a}, R. and {Sarasso}, M. and {Schultheis}, M. and {Sciacca}, E. and {Segol}, M. and {Segovia}, J.~C. and {S{\'e}gransan}, D. and {Semeux}, D. and {Shahaf}, S. and {Siddiqui}, H.~I. and {Siebert}, A. and {Siltala}, L. and {Slezak}, E. and {Smart}, R.~L. and {Solano}, E. and {Solitro}, F. and {Souami}, D. and {Souchay}, J. and {Spagna}, A. and {Spoto}, F. and {Steele}, I.~A. and {Steidelm{\"u}ller}, H. and {Stephenson}, C.~A. and {S{\"u}veges}, M. and {Szabados}, L. and {Szegedi-Elek}, E. and {Taris}, F. and {Tauran}, G. and {Taylor}, M.~B. and {Teixeira}, R. and {Thuillot}, W. and {Tonello}, N. and {Torra}, F. and {Torra}, J. and {Turon}, C. and {Unger}, N. and {Vaillant}, M. and {van Dillen}, E. and {Vanel}, O. and {Vecchiato}, A. and {Viala}, Y. and {Vicente}, D. and {Voutsinas}, S. and {Weiler}, M. and {Wevers}, T. and {Wyrzykowski}, {\L}. and {Yoldas}, A. and {Yvard}, P. and {Zhao}, H. and {Zorec}, J. and {Zucker}, S. and {Zurbach}, C. and {Zwitter}, T.},
        title = "{Gaia Early Data Release 3. Summary of the contents and survey properties}",
      journal = {\aap},
         year = 2021,
        month = may,
       volume = {649},
          eid = {A1},
        pages = {A1},
          doi = {10.1051/0004-6361/202039657},
archivePrefix = {arXiv},
       eprint = {2012.01533},
 primaryClass = {astro-ph.GA},
       adsurl = {https://ui.adsabs.harvard.edu/abs/2021A&A...649A...1G}
}

@BOOK{1957Zwicky,
       author = {{Zwicky}, Fritz},
        title = "{Morphological astronomy}",
         year = 1957,
       adsurl = {https://ui.adsabs.harvard.edu/abs/1957moas.book.....Z}
}

@ARTICLE{Blaauw1954,
       author = {{Blaauw}, A. and {Morgan}, W.~W.},
        title = "{The Space Motions of AE Aurigae and {\ensuremath{\mu}} Columbae with Respect to the Orion Nebula.}",
      journal = {\apj},
         year = 1954,
        month = may,
       volume = {119},
        pages = {625},
          doi = {10.1086/145866},
       adsurl = {https://ui.adsabs.harvard.edu/abs/1954ApJ...119..625B}
}

@ARTICLE{2001Hoogerwerf,
       author = {{Hoogerwerf}, R. and {de Bruijne}, J.~H.~J. and {de Zeeuw}, P.~T.},
        title = "{On the origin of the O and B-type stars with high velocities. II. Runaway stars and pulsars ejected from the nearby young stellar groups}",
      journal = {\aap},
         year = 2001,
        month = jan,
       volume = {365},
        pages = {49-77},
          doi = {10.1051/0004-6361:20000014},
archivePrefix = {arXiv},
       eprint = {astro-ph/0010057},
 primaryClass = {astro-ph},
       adsurl = {https://ui.adsabs.harvard.edu/abs/2001A&A...365...49H}
}

@ARTICLE{Hoogerwerf2000,
       author = {{Hoogerwerf}, R. and {de Bruijne}, J.~H.~J. and {de Zeeuw}, P.~T.},
        title = "{The Origin of Runaway Stars}",
      journal = {\apjl},
         year = 2000,
        month = dec,
       volume = {544},
       number = {2},
        pages = {L133-L136},
          doi = {10.1086/317315},
archivePrefix = {arXiv},
       eprint = {astro-ph/0007436},
 primaryClass = {astro-ph},
       adsurl = {https://ui.adsabs.harvard.edu/abs/2000ApJ...544L.133H}
}

@ARTICLE{2007LanztlustyB,
       author = {{Lanz}, Thierry and {Hubeny}, Ivan},
        title = "{A Grid of NLTE Line-blanketed Model Atmospheres of Early B-Type Stars}",
      journal = {\apjs},
         year = 2007,
        month = mar,
       volume = {169},
       number = {1},
        pages = {83-104},
          doi = {10.1086/511270},
archivePrefix = {arXiv},
       eprint = {astro-ph/0611891},
 primaryClass = {astro-ph},
       adsurl = {https://ui.adsabs.harvard.edu/abs/2007ApJS..169...83L}
}

@ARTICLE{1979Stone,
       author = {{Stone}, R.~C.},
        title = "{Kinematics, close binary evolution, and ages of the O stars.}",
      journal = {\apj},
         year = 1979,
        month = sep,
       volume = {232},
        pages = {520-530},
          doi = {10.1086/157311},
       adsurl = {https://ui.adsabs.harvard.edu/abs/1979ApJ...232..520S}
}

@ARTICLE{2024ApJSGYJ,
       author = {{Guo}, Yanjun and {Wang}, Luqian and {Liu}, Chao and {Wu}, You and {Han}, ZhanWen and {Chen}, XueFei},
        title = "{A Catalog of Early-type Runaway Stars from LAMOST DR8}",
      journal = {\apjs},
         year = 2024,
        month = jun,
       volume = {272},
       number = {2},
          eid = {45},
        pages = {45},
          doi = {10.3847/1538-4365/ad46f8},
archivePrefix = {arXiv},
       eprint = {2405.04750},
 primaryClass = {astro-ph.SR},
       adsurl = {https://ui.adsabs.harvard.edu/abs/2024ApJS..272...45G}
}

@ARTICLE{2012Bressan,
       author = {{Bressan}, Alessandro and {Marigo}, Paola and {Girardi}, L{\'e}o. and {Salasnich}, Bernardo and {Dal Cero}, Claudia and {Rubele}, Stefano and {Nanni}, Ambra},
        title = "{PARSEC: stellar tracks and isochrones with the PAdova and TRieste Stellar Evolution Code}",
      journal = {\mnras},
         year = 2012,
        month = nov,
       volume = {427},
       number = {1},
        pages = {127-145},
          doi = {10.1111/j.1365-2966.2012.21948.x},
archivePrefix = {arXiv},
       eprint = {1208.4498},
 primaryClass = {astro-ph.SR},
       adsurl = {https://ui.adsabs.harvard.edu/abs/2012MNRAS.427..127B}
}

@ARTICLE{2014Chen,
       author = {{Chen}, Yang and {Girardi}, L{\'e}o and {Bressan}, Alessandro and {Marigo}, Paola and {Barbieri}, Mauro and {Kong}, Xu},
        title = "{Improving PARSEC models for very low mass stars}",
      journal = {\mnras},
         year = 2014,
        month = nov,
       volume = {444},
       number = {3},
        pages = {2525-2543},
          doi = {10.1093/mnras/stu1605},
archivePrefix = {arXiv},
       eprint = {1409.0322},
 primaryClass = {astro-ph.SR},
       adsurl = {https://ui.adsabs.harvard.edu/abs/2014MNRAS.444.2525C}
}

@ARTICLE{2018Fu,
       author = {{Fu}, Xiaoting and {Bressan}, Alessandro and {Marigo}, Paola and {Girardi}, L{\'e}o and {Montalb{\'a}n}, Josefina and {Chen}, Yang and {Nanni}, Ambra},
        title = "{New PARSEC data base of {\ensuremath{\alpha}}-enhanced stellar evolutionary tracks and isochrones - I. Calibration with 47 Tuc (NGC 104) and the improvement on RGB bump}",
      journal = {\mnras},
         year = 2018,
        month = may,
       volume = {476},
       number = {1},
        pages = {496-511},
          doi = {10.1093/mnras/sty235},
archivePrefix = {arXiv},
       eprint = {1801.07137},
 primaryClass = {astro-ph.SR},
       adsurl = {https://ui.adsabs.harvard.edu/abs/2018MNRAS.476..496F}
}

@ARTICLE{2022Xiang,
       author = {{Xiang}, Maosheng and {Rix}, Hans-Walter and {Ting}, Yuan-Sen and {Kudritzki}, Rolf-Peter and {Conroy}, Charlie and {Zari}, Eleonora and {Shi}, Jian-Rong and {Przybilla}, Norbert and {Ramirez-Tannus}, Maria and {Tkachenko}, Andrew and {Gebruers}, Sarah and {Liu}, Xiao-Wei},
        title = "{Stellar labels for hot stars from low-resolution spectra. I. The HotPayne method and results for 330 000 stars from LAMOST DR6}",
      journal = {\aap},
         year = 2022,
        month = jun,
       volume = {662},
          eid = {A66},
        pages = {A66},
          doi = {10.1051/0004-6361/202141570},
archivePrefix = {arXiv},
       eprint = {2108.02878},
 primaryClass = {astro-ph.SR},
       adsurl = {https://ui.adsabs.harvard.edu/abs/2022A&A...662A..66X}
}

@ARTICLE{1995Herbig,
       author = {{Herbig}, G.~H.},
        title = "{The Diffuse Interstellar Bands}",
      journal = {\araa},
         year = 1995,
        month = jan,
       volume = {33},
        pages = {19-74},
          doi = {10.1146/annurev.aa.33.090195.000315},
       adsurl = {https://ui.adsabs.harvard.edu/abs/1995ARA&A..33...19H}
}

@ARTICLE{2021Gaiadis,
       author = {{Bailer-Jones}, C.~A.~L. and {Rybizki}, J. and {Fouesneau}, M. and {Demleitner}, M. and {Andrae}, R.},
        title = "{Estimating Distances from Parallaxes. V. Geometric and Photogeometric Distances to 1.47 Billion Stars in Gaia Early Data Release 3}",
      journal = {\aj},
         year = 2021,
        month = mar,
       volume = {161},
       number = {3},
          eid = {147},
        pages = {147},
          doi = {10.3847/1538-3881/abd806},
archivePrefix = {arXiv},
       eprint = {2012.05220},
 primaryClass = {astro-ph.SR},
       adsurl = {https://ui.adsabs.harvard.edu/abs/2021AJ....161..147B}
}

@ARTICLE{2015Bovygalpy,
       author = {{Bovy}, Jo},
        title = "{galpy: A python Library for Galactic Dynamics}",
      journal = {\apjs},
         year = 2015,
        month = feb,
       volume = {216},
       number = {2},
          eid = {29},
        pages = {29},
          doi = {10.1088/0067-0049/216/2/29},
archivePrefix = {arXiv},
       eprint = {1412.3451},
 primaryClass = {astro-ph.GA},
       adsurl = {https://ui.adsabs.harvard.edu/abs/2015ApJS..216...29B}
}

@ARTICLE{2023liuzhicunB,
       author = {{Liu}, Zhicun and {Cui}, Wenyuan and {Zhao}, Gang and {Liu}, Chao and {Luo}, Changqing and {Alexeeva}, Sofya},
        title = "{Origins of B-type stars at high Galactic latitudes based on abundances and kinematics}",
      journal = {\mnras},
         year = 2023,
        month = feb,
       volume = {519},
       number = {1},
        pages = {995-1012},
          doi = {10.1093/mnras/stac3562},
       adsurl = {https://ui.adsabs.harvard.edu/abs/2023MNRAS.519..995L}
}

@ARTICLE{2023Hunt,
       author = {{Hunt}, Emily L. and {Reffert}, Sabine},
        title = "{Improving the open cluster census. II. An all-sky cluster catalogue with Gaia DR3}",
      journal = {\aap},
         year = 2023,
        month = may,
       volume = {673},
          eid = {A114},
        pages = {A114},
          doi = {10.1051/0004-6361/202346285},
archivePrefix = {arXiv},
       eprint = {2303.13424},
 primaryClass = {astro-ph.GA},
       adsurl = {https://ui.adsabs.harvard.edu/abs/2023A&A...673A.114H}
}

@ARTICLE{2020Castro,
       author = {{Castro-Ginard}, A. and {Jordi}, C. and {Luri}, X. and {{\'A}lvarez Cid-Fuentes}, J. and {Casamiquela}, L. and {Anders}, F. and {Cantat-Gaudin}, T. and {Mongui{\'o}}, M. and {Balaguer-N{\'u}{\~n}ez}, L. and {Sol{\`a}}, S. and {Badia}, R.~M.},
        title = "{Hunting for open clusters in Gaia DR2: 582 new open clusters in the Galactic disc}",
      journal = {\aap},
         year = 2020,
        month = mar,
       volume = {635},
          eid = {A45},
        pages = {A45},
          doi = {10.1051/0004-6361/201937386},
archivePrefix = {arXiv},
       eprint = {2001.07122},
 primaryClass = {astro-ph.GA},
       adsurl = {https://ui.adsabs.harvard.edu/abs/2020A&A...635A..45C}
}

@ARTICLE{2019Castro,
       author = {{Castro-Ginard}, A. and {Jordi}, C. and {Luri}, X. and {Cantat-Gaudin}, T. and {Balaguer-N{\'u}{\~n}ez}, L.},
        title = "{Hunting for open clusters in Gaia DR2: the Galactic anticentre}",
      journal = {\aap},
         year = 2019,
        month = jul,
       volume = {627},
          eid = {A35},
        pages = {A35},
          doi = {10.1051/0004-6361/201935531},
archivePrefix = {arXiv},
       eprint = {1905.06161},
 primaryClass = {astro-ph.GA},
       adsurl = {https://ui.adsabs.harvard.edu/abs/2019A&A...627A..35C}
}

@PROCEEDINGS{1993Blaauw,
        title = "{Massive Stars: Their Lives in the Interstellar Medium}",
    booktitle = {Massive Stars:  Their Lives in the Interstellar Medium},
         year = 1993,
       editor = {{Cassinelli}, Joseph P. and {Churchwell}, Edward B.},
       volume = {35},
        month = jan,
       adsurl = {https://ui.adsabs.harvard.edu/abs/1993ASPC...35.....C}
}

@ARTICLE{2009Warren,
       author = {{Warren}, Steven R. and {Cole}, Andrew A.},
        title = "{Metallicities and radial velocities of five open clusters including a new candidate member of the Monoceros stream*}",
      journal = {\mnras},
         year = 2009,
        month = feb,
       volume = {393},
       number = {1},
        pages = {272-296},
          doi = {10.1111/j.1365-2966.2008.14268.x},
archivePrefix = {arXiv},
       eprint = {0811.2925},
 primaryClass = {astro-ph},
       adsurl = {https://ui.adsabs.harvard.edu/abs/2009MNRAS.393..272W}
}

@ARTICLE{2020Zhong,
       author = {{Zhong}, Jing and {Chen}, Li and {Wu}, Di and {Li}, Lu and {Bai}, Leya and {Hou}, Jinliang},
        title = "{Exploring open cluster properties with Gaia and LAMOST}",
      journal = {\aap},
         year = 2020,
        month = aug,
       volume = {640},
          eid = {A127},
        pages = {A127},
          doi = {10.1051/0004-6361/201937131},
archivePrefix = {arXiv},
       eprint = {2006.06929},
 primaryClass = {astro-ph.GA},
       adsurl = {https://ui.adsabs.harvard.edu/abs/2020A&A...640A.127Z}
}

@ARTICLE{2019Renzo,
       author = {{Renzo}, M. and {Zapartas}, E. and {de Mink}, S.~E. and {G{\"o}tberg}, Y. and {Justham}, S. and {Farmer}, R.~J. and {Izzard}, R.~G. and {Toonen}, S. and {Sana}, H.},
        title = "{Massive runaway and walkaway stars. A study of the kinematical imprints of the physical processes governing the evolution and explosion of their binary progenitors}",
      journal = {\aap},
         year = 2019,
        month = apr,
       volume = {624},
          eid = {A66},
        pages = {A66},
          doi = {10.1051/0004-6361/201833297},
archivePrefix = {arXiv},
       eprint = {1804.09164},
 primaryClass = {astro-ph.SR},
       adsurl = {https://ui.adsabs.harvard.edu/abs/2019A&A...624A..66R}
}

@ARTICLE{2000van,
       author = {{van den Heuvel}, E.~P.~J. and {Portegies Zwart}, S.~F. and {Bhattacharya}, D. and {Kaper}, L.},
        title = "{On the origin of the difference between the runaway velocities of the OB-supergiant X-ray binaries and the Be/X-ray binaries}",
      journal = {\aap},
         year = 2000,
        month = dec,
       volume = {364},
        pages = {563-572},
          doi = {10.48550/arXiv.astro-ph/0005245},
archivePrefix = {arXiv},
       eprint = {astro-ph/0005245},
 primaryClass = {astro-ph},
       adsurl = {https://ui.adsabs.harvard.edu/abs/2000A&A...364..563V}
}

@ARTICLE{1983Hoffer,
       author = {{Hoffer}, J.~B.},
        title = "{Computer simulations of gravitational encounters between pairs of binary star systems.}",
      journal = {\aj},
         year = 1983,
        month = sep,
       volume = {88},
        pages = {1420-1434},
          doi = {10.1086/113431},
       adsurl = {https://ui.adsabs.harvard.edu/abs/1983AJ.....88.1420H}
}

@ARTICLE{2025CK,
       author = {{Chen}, Kun and {Guo}, Yanjun and {Jiang}, Dengkai and {Han}, Zhanwen and {Chen}, Xuefei},
        title = "{The Binary Fraction of B-type Runaway Stars from LAMOST DR8}",
      journal = {\apj},
         year = 2025,
        month = aug,
       volume = {988},
       number = {2},
          eid = {228},
        pages = {228},
          doi = {10.3847/1538-4357/ade702},
       adsurl = {https://ui.adsabs.harvard.edu/abs/2025ApJ...988..228C}
}

@ARTICLE{2017McEvoy,
       author = {{McEvoy}, Catherine M. and {Dufton}, Philip L. and {Smoker}, Jonathan V. and {Lambert}, David L. and {Keenan}, Francis P. and {Schneider}, Fabian R.~N. and {de Wit}, Willem-Jan},
        title = "{The Origin of B-type Runaway Stars: Non-LTE Abundances as a Diagnostic}",
      journal = {\apj},
         year = 2017,
        month = jun,
       volume = {842},
       number = {1},
          eid = {32},
        pages = {32},
          doi = {10.3847/1538-4357/aa745a},
archivePrefix = {arXiv},
       eprint = {1708.03527},
 primaryClass = {astro-ph.SR},
       adsurl = {https://ui.adsabs.harvard.edu/abs/2017ApJ...842...32M}
}

@ARTICLE{2008Przybilla,
       author = {{Przybilla}, Norbert and {Fernanda Nieva}, M. and {Heber}, Ulrich and {Butler}, Keith},
        title = "{HD 271791: An Extreme Supernova Runaway B Star Escaping from the Galaxy}",
      journal = {\apjl},
         year = 2008,
        month = sep,
       volume = {684},
       number = {2},
        pages = {L103},
          doi = {10.1086/592245},
archivePrefix = {arXiv},
       eprint = {0811.0576},
 primaryClass = {astro-ph},
       adsurl = {https://ui.adsabs.harvard.edu/abs/2008ApJ...684L.103P}
}

@ARTICLE{2010Irrgang,
       author = {{Irrgang}, Andreas and {Przybilla}, Norbert and {Heber}, Ulrich and {Nieva}, M. Fernanda and {Schuh}, Sonja},
        title = "{The Nature of the Hyper-Runaway Candidate Hip 60350}",
      journal = {\apj},
         year = 2010,
        month = mar,
       volume = {711},
       number = {1},
        pages = {138-143},
          doi = {10.1088/0004-637X/711/1/138},
archivePrefix = {arXiv},
       eprint = {1002.1848},
 primaryClass = {astro-ph.SR},
       adsurl = {https://ui.adsabs.harvard.edu/abs/2010ApJ...711..138I}
}

\begin{appendix}
\section{Table}\label{app:Table}
\begin{sidewaystable*}
\caption{\label{Tab:1}The predicted parameters of B-type runaway stars.}
\centering
\scriptsize
\setlength{\tabcolsep}{3pt}
\renewcommand{\arraystretch}{1.1}
\begin{tabular}{lccccccccccccccccccccccccl}
 \hline \hline
 \GG{Gaia DR3} ID& Name &RA    & DEC  & Para & $\mu_\alpha$ & $\mu_\beta$ &   $T_\mathrm{eff}$   & $\log{g}$ & $[M/H]$  & $v\sin{i}$  & Mass &  Age & $V_{Sp}$ & $X$ & $Y$ & $Z$ & $U$ & $V$ & $W$ & Class & Flag\\  
&&(deg) & (deg) & (mas) & (mas yr$^{-1}$) & (mas yr$^{-1}$) & (K)  &  (dex)   &     (dex)         & (km s$^{-1}$) & (${\rm M}_{\sun}$) & (Myr) & (km s$^{-1}$) & (kpc) & (kpc)& (kpc) & (km s$^{-1}$)&(km s$^{-1}$)&(km s$^{-1}$)& &\\ \hline
\G{454303606011095552 }&\G{LS V+545       }  & 41.4488  & 54.5447  &\G{0.3927} &\G{0.803 }&\G{-2.306} 	& 25621  & 4.1  & $-$0.2 & 29  & 10.0$\pm$ 1.2 & 12.4$\pm$ 4.7  &\G{34.24}  &\G{10.19}  &\G{1.66 } &\G{0.23}  & \G{11.32 }  &\G{194.09}  &\G{-14.63} & B2  & I \\
\G{160785506636872064 }&\G{HD 282581      }  & 74.1620  & 32.9081  &\G{0.733 } &\G{1.555 }&\G{-6.167}	& 15841  & 3.6  & $-$0.8 & 158  & 5.3$\pm$ 0.6 & 78.1$\pm$ 30.6 &\G{27.8 }  &\G{9.61 }  &\G{0.22 } &\G{0.17}  & \G{4.78  }  &\G{192.62}  &\G{3.64 } & B5   & I \\
\G{3443256628164281216}&\G{               }  & 87.9204  & 28.9357  &\G{0.285 } &\G{1.952 }&\G{-2.237} 	& 16643  & 4.0  & $-$0.5 & 78  & 4.8$\pm$ 0.6 & 79.2$\pm$ 25.5  &\G{41.01}  &\G{11.78}  &\G{-0.04} &\G{0.09}  & \G{-40.07}  &\G{241.19}  &\G{10.33 }& B5   & I \\
\G{3326494643686889088}&\G{TYC  733-1378-1}  & 99.5149  & 8.6713   &\G{0.6573} &\G{-1.109}&\G{2.748	}   & 14466  & 4.0  & $-$1.3 & 0  & 3.1$\pm$ 0.4 & 271.3$\pm$ 93.1  &\G{29.29}  &\G{9.66 }  &\G{-0.61} &\G{0.05}  & \G{-8.21 }  &\G{216.6 }  &\G{-18.13}& B7   & I \\
\G{3326494643686889088}&\G{LS  VI +00 39  }  & 106.4899 & 0.4331   &\G{0.1677} &\G{0.57  }&\G{0.564 }   & 18337  & 3.5  & $-$0.5 & 0  & 8$\pm$ 1.5 & 36.1$\pm$ 13.9     &\G{46.17}  &\G{13.19}  &\G{-3.35} &\G{0.38}  & \G{-68.24}  &\G{186.58}  &\G{34.37 }& B2-4 & I \\
\hline  \hline
\end{tabular}
\end{sidewaystable*}
\end{appendix}

\end{document}